\documentclass[5p]{elsarticle}

\makeatletter
\def\ps@pprintTitle{%
 \let\@oddhead\@empty
 \let\@evenhead\@empty
 \def\@oddfoot{\footnotesize\itshape
       Preprint submitted to \ifx\@journal\@empty Elsevier
       \else\@journal\fi\hfill March 09, 2019}%
 \let\@evenfoot\@oddfoot}
\makeatother

\usepackage{amsmath}
\usepackage{amsfonts}
\usepackage{amssymb}
\usepackage{bm}
\usepackage{undertilde}
\usepackage{xspace}
\usepackage{siunitx}

\usepackage[colorlinks]{hyperref}

\journal{Ocean Modelling}

\biboptions{authoryear}

\newcommand{\di}{\partial}

\newcommand{\ppx}[1]{\frac{\partial}{\partial{#1}}}

\newcommand{\dydx}[2]{\frac{\mathrm{d}{#1}}{\mathrm{d}{#2}}}

\newcommand{\ie}{i.e.\ }
\newcommand{\eg}{e.g.\ }

\newcommand{\vs}{vs.\ }

\newcommand{\ihat}{\bm{\hat{i}}}
\newcommand{\jhat}{\bm{\hat{j}}}
\newcommand{\khat}{\bm{\hat{k}}}

\newcommand{\rB}{\rho_B} 

\newcommand{\eos}{R}
\newcommand{\eosB}{B}
\newcommand{\pn}{\utilde{p}}
\newcommand{\zn}{\utilde{z}}
\newcommand{\sn}{\utilde{S}}
\newcommand{\tn}{\utilde{\theta}}
\newcommand{\rn}{\utilde{\rho}}
\newcommand{\rpn}{\utilde{\rho_p}}

\newcommand{\rsn}{\utilde{\rho_S}}
\newcommand{\rtn}{\utilde{\rho_\theta}}
\newcommand{\rfn}{\hat{\rho}} 
\newcommand{\rpfn}{\hat{\pi}} 
\newcommand{\rzfn}{\hat{\pi}} 

\newcommand{\M}{\mathbb{M}} 

\newcommand{\rv}{0} 
\newcommand{\parg}{\mathtt{p}} 
\newcommand{\reg}{\mathcal{R}} 
\newcommand{\node}[1]{\textsf{#1}} 
\newcommand{\lon}{\lambda}
\newcommand{\lat}{\phi}

\newcommand{\textdef}{\textbf}

\newcommand{\xref}{180^\circ\mathrm{E}}
\newcommand{\yref}{0^\circ\mathrm{N}}
\newcommand{\zrefone}{-997.99}
\newcommand{\zreftwo}{-1988.60}

\begin{document}

\begin{frontmatter}

\title{Neutral surface topology}

\author{Geoffrey J. Stanley\fnref{fn1}}
\address{Department of Physics, University of Oxford, Oxford, OX1 3PU, United Kingdom}
\ead{g.stanley@unsw.edu.au}

\fntext[fn1]{Current address: School of Mathematics and Statistics, University of New South Wales, Sydney, NSW 2052, Australia. \\
\copyright{} 2019. This manuscript version is made available under the CC-BY-NC-ND 4.0 license http://creativecommons.org/licenses/by-nc-nd/4.0/}


\begin{abstract}
Neutral surfaces, along which most of the mixing in the ocean occurs, are notoriously difficult objects: they do not exist as well-defined surfaces, and as such can only be approximated. In a hypothetical ocean where neutral surfaces are well-defined, the \emph{in-situ} density on the surface is a multivalued function of the pressure on the surface, $\utilde{p}$. The surface is decomposed into geographic regions where there is one connected pressure contour per pressure value, making this function single-valued in each region. The regions are represented by arcs of the Reeb graph of $\utilde{p}$. The regions meet at saddles of $\utilde{p}$ which are represented by internal nodes of the Reeb graph. Leaf nodes represent extrema of $\utilde{p}$. Cycles in the Reeb graph are created by islands and other holes in the neutral surface. This topological theory of neutral surfaces is used to create a new class of approximately neutral surfaces in the real ocean, called topobaric surfaces, which are very close to neutral and fast to compute. Topobaric surfaces are the topologically correct extension of orthobaric density surfaces to be geographically dependent, which is fundamental to neutral surfaces. Also considered is the possibility that helical neutral trajectories might have a larger pitch around islands than in the open ocean.
\end{abstract}


\end{frontmatter}


\section{Introduction}
\label{sec:intro}

Strong stratification throughout most of the ocean inhibits vertical motion, largely confining the oceanic flow to a two-dimensional surface, the ideal of which is called a neutral surface \citep{mcdougall1987ns}. These surfaces are far from flat, and it is along these sloping surfaces that oceanic flows efficiently mix tracers (epineutral mixing), whereas tracer mixing across them (dianeutral mixing) is enormously weaker---an idea tracing back to \citet{iselin1939}. This is a great conceptual simplification, but only useful if we can map the depth, or pressure, of such surfaces. 
Unfortunately, non-linearity in seawater's equation of state leads to a path-dependence underlying the definition of neutral surfaces, making neutral surfaces ill-defined \citep{mcdougall.jackett1988}.

Given this difficulty, physical oceanographers craft well-defined surfaces that are everywhere nearly tangent to the neutral tangent plane, called approximately neutral surfaces. These surfaces are usually isosurfaces of a 3D variable, the earliest being potential density \citep{wust1935} and specific volume anomaly \citep{montgomery1937}.  \citet{lynn.reid1968} revealed the highly undesirable property that, far away from its reference pressure, potential density surfaces (isopycnals) in a stably stratified ocean can exhibit unphysical overturns. This problem also affects specific volume anomaly surfaces, far from the reference values. 
 
To overcome this, \citet{reid.lynn1971} introduced patched potential density.
They map the $\sigma_4 = 45.92$ potential density surface (referenced to \SI{4000}{dbar}) in the tropical Atlantic, and where this surface rises above \SI{-3000}{m} in the North Atlantic, it is patched together with the $\sigma_2 = 37.14$ potential density surface (referenced to \SI{2000}{dbar}). In fact, $\sigma_2$ varies somewhat along the length of this \SI{-3000}{m} contour, and 37.14 is chosen to minimize this discontinuity. 
Similarly, where the $\sigma_4 = 45.92$ surface rises above \SI{-3000}{m} in the South Atlantic, they patch it together with the $\sigma_2 = 37.10$ surface.
Noting that a single $\sigma_4$ surface is patched together with different $\sigma_2$ surfaces in the North Atlantic and Southern Ocean, it is clear that neutral surfaces are not just dependent upon salinity, temperature, and pressure, but also upon geography (latitude and longitude). 

In this way, the ocean may be cut, stacked, and arranged into boxes covering certain depth ranges and horizontal areas.
Where to make these cuts is not entirely arbitrary. Indeed, the equatorial Atlantic is a good place to make a cut, where a ``bridge'' region connects saltier North Atlantic waters with fresher South Atlantic waters; salinity and potential temperature are tightly related in these three regions, but the functional relationship differs between regions \citep{deszoeke.springer2005}. 

Though not entirely arbitrary, these divisions are also not entirely correct. There is nothing particularly important, thermodynamically, about the equator, nor about any other latitude circle, longitude circle, or depth level. There \emph{are} thermodynamically important regions, but their shape is not so simple. To study their shape is to study topology. 

This paper presents a fresh theoretical perspective on this problem, by studying the topology of hypothetical neutral surfaces that are well-defined.
On such surfaces, there is a multivalued functional relationship between the \emph{in-situ} density and the pressure. 
Different branches of this multivalued function arise because a level set of pressure on a neutral surface can be the disjoint union of multiple connected components, each of which supports a distinct \emph{in-situ} density. 
The important topological information about changes in the connectedness of these level sets is captured, as a collection of nodes and arcs, by the \citet{reeb1946} graph. This also determines the shape of the regions inside which the aforementioned multivalued function actually is just single-valued. 
This theoretical tool is then used to develop a new class of approximately neutral surfaces in the real ocean, called topobaric surfaces. 
Topobaric surfaces are very close to neutral and possess an exact geostrophic streamfunction \citep{stanley2019geostrf}.
Moreover, they are fast to compute: Computational topology is a young field, but efficient algorithms to compute the Reeb graph have recently been developed \citep{doraiswamy.natarajan2013}.

Though defined over 70 years ago, the Reeb graph has not, to the best of the author's knowledge, been previously used in oceanography, nor as a way of studying multivalued functional relationships between variables. 
The Reeb graph was most famously used by \citet{arnold1957} in solving Hilbert's superposition problem \citep[see also][]{arnold2006}, but its primary use of late is in computer graphics and visualization \citep[see][for a review]{biasotti.giorgi.ea2008}.

The paper is structured as follows.
The theory of neutral surfaces is reviewed in Section~\ref{sec:background}, then developed in Section~\ref{sec:topology_neutral_surfaces} from a topological perspective, discussing the Reeb graph, as well as the role of islands in making neutral surfaces ill-defined. 
A pedagogical illustration of the multivalued functional relationship and the Reeb graph is given in Section~\ref{sec:illustrative}. 
Section~\ref{sec:topobaric} discusses topobaric surfaces, from their theoretical description to their numerical calculation, and finally to their evaluation as useful approximately neutral surfaces. 
Conclusions are given in Section~\ref{sec:conclusions}.
A glossary of graph theory definitions is given in \ref{sec:mathdefs_graphtheory}, and 
\ref{sec:island_pitch} contains a preliminary analysis of the role of islands in helical neutral motions.

\section{Background of neutral surfaces}
\label{sec:background}

\subsection{Definitions}

In the ocean, the salinity $S$, potential temperature $\theta$, and pressure $p$ determine the \emph{in-situ} density $\rho$ according to a function $\eos$, the equation of state.\footnote{
Everything could be described in terms of Absolute Salinity and Conservative Temperature, and this would be completely equivalent for the theory presented here. This paper speaks of practical salinity and potential temperature simply because these are the inputs to the \citet{jackett.mcdougall1995} equation of state that is used by the ocean model whose data shall be analysed. 
} Mathematically, $\rho = \eos(S,\theta,p)$, where $S$, $\theta$, $p$, and $\rho$ are 3D scalar fields.
Using the chain rule, the gradient of \emph{in-situ} density is 
\begin{equation}
\label{eq:gradsv}
\nabla \rho = \rho_S \nabla S + \rho_\theta \nabla \theta + \rho_p \nabla p,
\end{equation}
where $\rho_S = \di_S \eos (S, \theta, p)$, $\rho_\theta = \di_\theta \eos(S, \theta, p)$, and $\rho_p = \di_p \eos(S, \theta, p)$ are 3D scalar fields.\footnote{
Often \eqref{eq:gradsv} is divided by $\rho$, so as to use the thermal expansion coefficient $-\rho^{-1} \rho_\theta$,
the haline contraction coefficient $\rho^{-1} \rho_S$, 
and the adiabatic compressibility $\rho^{-1} \rho_p$. 
This complicates further differentiation and so is not used here. 
}
Let $\ihat$, $\jhat$, and $\khat$ be the eastward, northward, and radial (vertical) unit vectors, respectively.

Consider displacing a fluid parcel of \emph{in-situ} density $\rho_0$ infinitesimally by $\mathrm{d} \bm{r}$. 
Its new surroundings have \emph{in-situ} density $\rho_0 + \nabla \rho \cdot \mathrm{d} \bm{r}$. Using \eqref{eq:gradsv}, this has contributions from salt, potential temperature, and pressure changes.
If the fluid parcel is insulated (meaning the salinity and potential temperature are conserved, commonly called adiabatic), its \emph{in-situ} density after displacement is only $\rho_0 + \rho_p \nabla p \cdot \mathrm{d} \bm{r}$. The difference between the environmental and parcel \emph{in-situ} density, $(\rho_S \nabla S + \rho_\theta \nabla \theta) \cdot \mathrm{d} \bm{r}$, creates a buoyant restoring force. Thus, if the displacement $\mathrm{d} \bm{r}$ is perpendicular to the \textdef{dianeutral vector}
\begin{equation}
\label{eq:def:dianeutral_vector}
\bm{N} = \rho_S \nabla S + \rho_\theta \nabla \theta,
\end{equation}
then the buoyant restoring force is zero. 
The \textdef{neutral tangent plane} is the plane normal to the dianeutral vector \citep{mcdougall1987ns}. A fluid parcel can move infinitesimally in this plane without experiencing a buoyant restoring force.\footnote{
Noting that mixing along the neutral tangent plane does require some kinetic energy, \citet{nycander2011} defined a vector $\bm{P}$ using the dynamic enthalpy rather than the equation of state. The form of $\bm{P}$ is the same as $\bm{N}$, so the essential ideas of this manuscript will apply equally well to surfaces formed from $\bm{N}$ or $\bm{P}$. 
}

A \textdef{neutral trajectory} is a solution of the Pfaffian differential equation $\bm{N} \cdot \, \mathrm{d} \bm{r} = 0$. That is, a neutral trajectory is a path that is always orthogonal to $\bm{N}$ and hence always tangent to the local neutral tangent plane. 

A \textdef{neutral surface} is a surface that is everywhere tangent to the neutral tangent plane. 

\subsection{Non-existence of neutral surfaces}
\label{sec:nonexistence}

Following \citet{mcdougall.jackett1988}, 
suppose we have a neutral surface that is a well-defined, 2D surface.  Consider a subset $\Omega$ of the neutral surface, having boundary $\di \Omega$, which is a neutral trajectory. 
The path integral of $\bm{N}$ along $\di \Omega$ is related to an area integral of the curl of $\bm{N}$ via Stokes' theorem: 
\begin{equation}
\label{eq:intA}
\int_{\di \Omega} \bm{N} \cdot \mathrm{d} \bm{r} = \int_\Omega \nabla \times \bm{N} \cdot \mathrm{d} \bm{\Omega}.
\end{equation}
The LHS is zero by the definition of a neutral trajectory. 
For the RHS, the surface of integration is a neutral surface so $\mathrm{d} \bm{\Omega} = \mathrm{d} \Omega \,\bm{N} / |\bm{N}|$, and hence 
\begin{equation}
\label{eq:intH}
0 = \int_\Omega \frac{H}{|\bm{N}|} \ \mathrm{d} \Omega,
\end{equation}
where $H = \bm{N} \cdot \nabla \times \bm{N}$ is the \textdef{neutral helicity}.
As $\Omega$ is arbitrary, it must be that $H = 0$ everywhere. 
Expanding $\bm{N}$ and using the chain rule on $\nabla \rho_S$ and $\nabla \rho_\theta$ akin to \eqref{eq:gradsv}, we find 
\begin{equation}
\label{eq:helicity}
H = (\rho_\theta \rho_{S p} - \rho_S \rho_{\theta p}) \nabla p \cdot \nabla S \times \nabla \theta,
\end{equation}
where $\rho_{S p} = \di_{S p} \eos(S,\theta,p)$ and $\rho_{\theta p} = \di_{\theta p} \eos(S,\theta,p)$ are 3D fields.
With a non-linear equation of state $\eos$, the term in parentheses is non-zero; it is $(\rho_S \rho)$ times the thermobaricity \citep[defined by][]{mcdougall1987tb}.
Thus $H = 0$ if and only if the three vectors $\nabla p$, $\nabla S$, and $\nabla \theta$ are coplanar. 
As this is not generally true \citep[see][for analysis of oceanic data]{mcdougall.jackett2007}, 
a contradiction is reached. 
The false assumption was to assume the neutral surface was a well-defined, 2D surface. 
Even if the ocean is stably stratified ($\khat \cdot \bm{N} < 0$ everywhere), the neutral trajectory $\di \Omega$ has returned to its initial geographic location but at a different depth from which it began. 
This phenomenon is called the \textdef{neutral helix}, and this depth change is called the \textdef{pitch} \citep{mcdougall.jackett1988}.

A neutral helix can be arbitrarily shrunk, in terms of its radially projected area, to produce a new neutral helix with a different pitch.\footnote{If the ocean is neutrally stable somewhere, the neutral tangent plane there contains the radial (vertical) direction. A better version of this argument is to measure the pitch in the $\bm{N}$ direction.}
This new pitch is a continuous function of the factor by which this area is shrunk, so any desired pitch, between the original pitch and zero, can be found: this is guaranteed by the intermediate value theorem. Thus, neutral trajectories from a given point are a set of points that extend laterally, along the neutral tangent planes, as well as radially (vertically), and so occupy a 3D volume rather than a 2D surface. 
This is how a neutral surface in an $H = 0$ ocean fails to be a well-defined, 2D surface.

\subsection{Well-defined surfaces \vs unique-depth surfaces}

We have just seen that a necessary condition for a neutral surface to be a well-defined surface is that $H = 0$ everywhere on that surface. 
\citet{jackett.mcdougall1997} asserted that $H = 0$ globally is also a sufficient condition to ensure neutral surfaces are well-defined surfaces, and this thinking has persisted \citep{mcdougall.jackett2007}.

This is correct, if one takes a well-defined surface by the standard mathematical and topological definition: a \textdef{well-defined surface} is a 2-manifold. 
Roughly speaking, this means it locally resembles 2D Euclidean space: every point on a 2-manifold has a neighbourhood that can be continuously deformed into an open subset of $\mathbb{R}^2$, and back again. 
In a stably stratified ocean, however, we would ideally like neutral surfaces to be \textdef{unique-depth surfaces}: well-defined surfaces whose depth is a single-valued function of geographic location (though undefined where the surface has grounded or outcropped). 

In fact, $H = 0$ globally is not sufficient to ensure neutral surfaces are unique-depth surfaces in a stably stratified ocean.
The aforementioned assertion by \citet{jackett.mcdougall1997} derives from the classic result \citep{sneddon1957}
 that the Pfaffian differential equation $\bm{N} \cdot \mathrm{d} \bm{r} = 0$ (solutions of which are neutral trajectories) is integrable if and only if $H = \bm{N} \cdot \nabla \times \bm{N} = 0$. However, this classic result holds only in an infinite domain,
not in a domain like the real ocean that
is bounded and replete with holes such as islands.

Islands and other such holes are important because neutral helices can exist around them even when $H = 0$ everywhere in the ocean.
A quick way to see this is to imagine the ocean with $H = 0$ everywhere except for some region, then build an artificial island over that region (taking great care to not otherwise disturb the ocean state): helical neutral trajectories that existed before the island construction still exist after it, but now $H = 0$ everywhere.

\begin{figure}[tb]
  \centering\includegraphics[width=1\columnwidth]{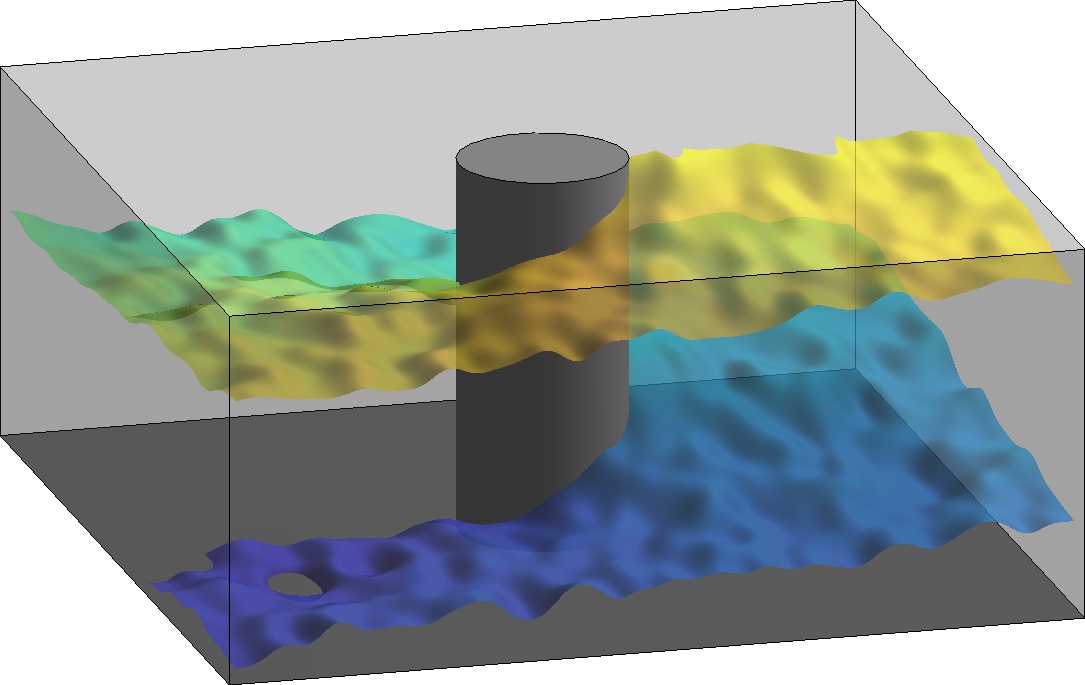}
  \caption{Schematic of a perfectly neutral surface (colours corresponding to its depth) around an island in a bounded ocean. Though the neutral helicity may be zero in this ocean, the island can effectively ``hide'' some non-zero neutral helicity, so the neutral surface can intersect some water columns at multiple depths: it is not a ``unique-depth surface''.
  }
\label{fig:island_helix}
\end{figure}

In this case, neutral helices exist but only around islands, so cannot be arbitrarily resized (as in Section~\ref{sec:nonexistence}). Each helix has a definite, non-infinitesimal, pitch. In such an $H = 0$ ocean, neutral surfaces are well-defined surfaces but not unique-depth surfaces. They resemble a multistorey car park with an interior ramp, as schematised in Fig.~\ref{fig:island_helix}. With many holes in the neutral surface, there may be many interior ramps. 
Note that holes in a neutral surface are created not just by islands, but also possibly where the surface outcrops or grounds, even in a flat-bottomed ocean, such as in the bottom-left of Fig.~\ref{fig:island_helix}.
However, a hole in the surface does not necessarily produce a neutral helix, a point we shall return to.

\subsection{Gradients on the sphere}
\label{sec:projnonorthgrad}

A neutral surface is supposed to be a surface that is everywhere tangent to the neutral tangent plane. To mathematise this, the 3D gradient of a tracer $C$ is decomposed into components parallel and normal to the neutral tangent plane, as such:
$\nabla C = \nabla_N C + (\nabla C \cdot \hat{\bm{n}}) \hat{\bm{n}}$ 
where $\hat{\bm{n}} = \bm{N} / |\bm{N}|$ \citep{mcdougall.groeskamp.ea2014}. 
Thus, displacements $\mathrm{d} \bm{r}$ in the neutral tangent plane satisfy
\begin{equation}
\label{eq:def_ntp_A}
0 = \mathrm{d} \bm{r} \cdot \bm{N}
= \mathrm{d} \bm{r} \cdot (\rho_S \nabla_N S + \rho_\theta \nabla_N \theta),
\end{equation}
having used $\mathrm{d} \bm{r} \cdot \hat{\bm{n}} = 0$ and \eqref{eq:def:dianeutral_vector}.
Since $(\rho_S \nabla_N S + \rho_\theta \nabla_N \theta)$ is in the neutral tangent plane and $\mathrm{d} \bm{r}$ is arbitrary in that plane, then $\hat{\bm{n}}$ must be such that
\begin{equation}
\label{eq:def_ntp_st}
\rho_S \nabla_N S + \rho_\theta \nabla_N \theta = \bm{0}.
\end{equation}

Whereas $\nabla_N C$ is a 3D vector field, it is useful to instead work with a 2D vector field determined only from quantities in the surface. To this end, we use the ``projected non-orthogonal gradient'' introduced by \citet{starr1945}. This is used throughout studies of neutral surfaces and commonly denoted $\nabla_n \equiv \ppx{x}\vert_n  \ihat + \ppx{y}\vert_n \jhat$. Applying this gradient to a 3D scalar field $C$ produces a 2D vector field, \ie $\nabla_n C \cdot \khat = 0$. The partial derivatives in $\nabla_n$ sample $C$ from the surface in question, but distances are measured only by their horizontal contribution (\ie they are projected onto a constant height).

This paper uses $\nabla \utilde{C}$ as an alternative notation for $\nabla_n C$. The \textdef{under-tilde} is an operator that restricts a 3D field to the surface in question, then projects it onto a sphere (sharing Earth's centre and radius). The result of this projection, denoted $\M$, is a subset of the sphere, with holes where the Earth has islands and continents, and where the original surface had grounded or outcropped.
For example, $\pn : \M \rightarrow \mathbb{R}$ is a scalar field (with physical units implied) on $\M$; using a geographic coordinate system, $\pn(\lon,\lat)$ is the pressure at longitude $\lon$ and latitude $\lat$ on the original surface in question. Hence, $\pn$ will be loosely referred to as the pressure on the surface in question, even though $\pn$ lives on $\M$. 
The gradient $\nabla \pn$ is a standard gradient, calculable in spherical coordinates: distance is measured without regard to the radial (vertical) variations of the original surface in question.
When the surface in question is a neutral surface, $\nabla \utilde{C}$ and $\nabla_n C$ differ only in that the former specifies a single surface at a time, whereas the latter is a 2D vector field living in 3D space. 

\citet{mcdougall.groeskamp.ea2014} showed that \eqref{eq:def_ntp_st} is equivalently expressed using the projected non-orthogonal gradient (simply change ``N'' to ``n''). With the under-tilde notation, this gives
\begin{equation}
\label{eq:def_ntp_st2}
\utilde{\rho_S} \nabla \sn + \utilde{\rho_\theta} \nabla \tn = \bm{0}.
\end{equation} 
A truly neutral surface must satisfy \eqref{eq:def_ntp_st2} exactly (though this is impossible in the real ocean with $H \neq 0$). 
Indeed, \eqref{eq:def_ntp_st2} is identical to the first of two equivalent definitions of neutral surfaces given by \citet{mcdougall1987ns}, the other being identical to
\begin{equation}
\label{eq:def_ntp_svp}
\nabla \rn =  \rpn \nabla \pn,
\end{equation}
which is derived like \eqref{eq:def_ntp_st2} but using \eqref{eq:gradsv} to express $\bm{N} = \nabla \rho - \rho_p \nabla p$.  
Defining neutral surfaces by \eqref{eq:def_ntp_svp}, rather than \eqref{eq:def_ntp_st2}, is preferable in this work because pressure is monotonic with depth.

\section{Neutral surface topology}
\label{sec:topology_neutral_surfaces}

In this section, we study the topology of properties on a well-defined neutral surface in a hypothetical ocean in which the neutral helicity is everywhere zero.\footnote{
If the surface is not unique-depth, $\M$ can have multiple ``layers'', but the following theory works equally well. Alternatively, one could work on the neutral surface itself rather than projecting it to the sphere; then $\M$ is a more general Riemannian manifold, and gradients reformulated in terms of the exterior derivative. This is not pursued here, for pedagogical and practical reasons.}
In this ideal setting, the exactness of \eqref{eq:def_ntp_svp} has global topological implications. This insight will be used to form topobaric surfaces (Section~\ref{sec:topobaric}), which are approximately neutral surfaces in the real ocean with non-zero neutral helicity.

First, we must distinguish between contours and level sets. A level set is the disjoint union of any number of contours. They are defined mathematically as follows.

A \textdef{path} from $\bm{x} \in \M$ to $\bm{y} \in \M$ is a continuous function $\mathcal{P} : [0,1] \rightarrow \M$ having $\mathcal{P}(0) = \bm{x}$, and $\mathcal{P}(1) = \bm{y}$.

The \textdef{contour} of $\pn$ through a point $\bm{x} \in \M$, denoted $\pn^{-1}(\bm{x})$, is the set of all $\bm{y} \in \M$ for which there exists a path $\mathcal{P}$ from $\bm{x}$ to $\bm{y}$ having $\pn(\mathcal{P}(t)) = \pn(\bm{x})$ for all $t \in [0, 1]$.

The \textdef{level set} of $\pn$ at $p' \in \mathbb{R}$ is the set of all $\bm{x} \in \M$ such that $\pn(\bm{x}) = p'$. Mathematically, it is $\pn^{-1}(p') = \{ \bm{x} \in \M : \pn(\bm{x}) = p' \}$.  The level set of a 3D field is often called an isosurface. 

\subsection{Single-valued functional relations}
\label{sec:singlevalued_functional_relations}

\begin{figure}[tb]
  \centering\includegraphics[width=1\columnwidth]{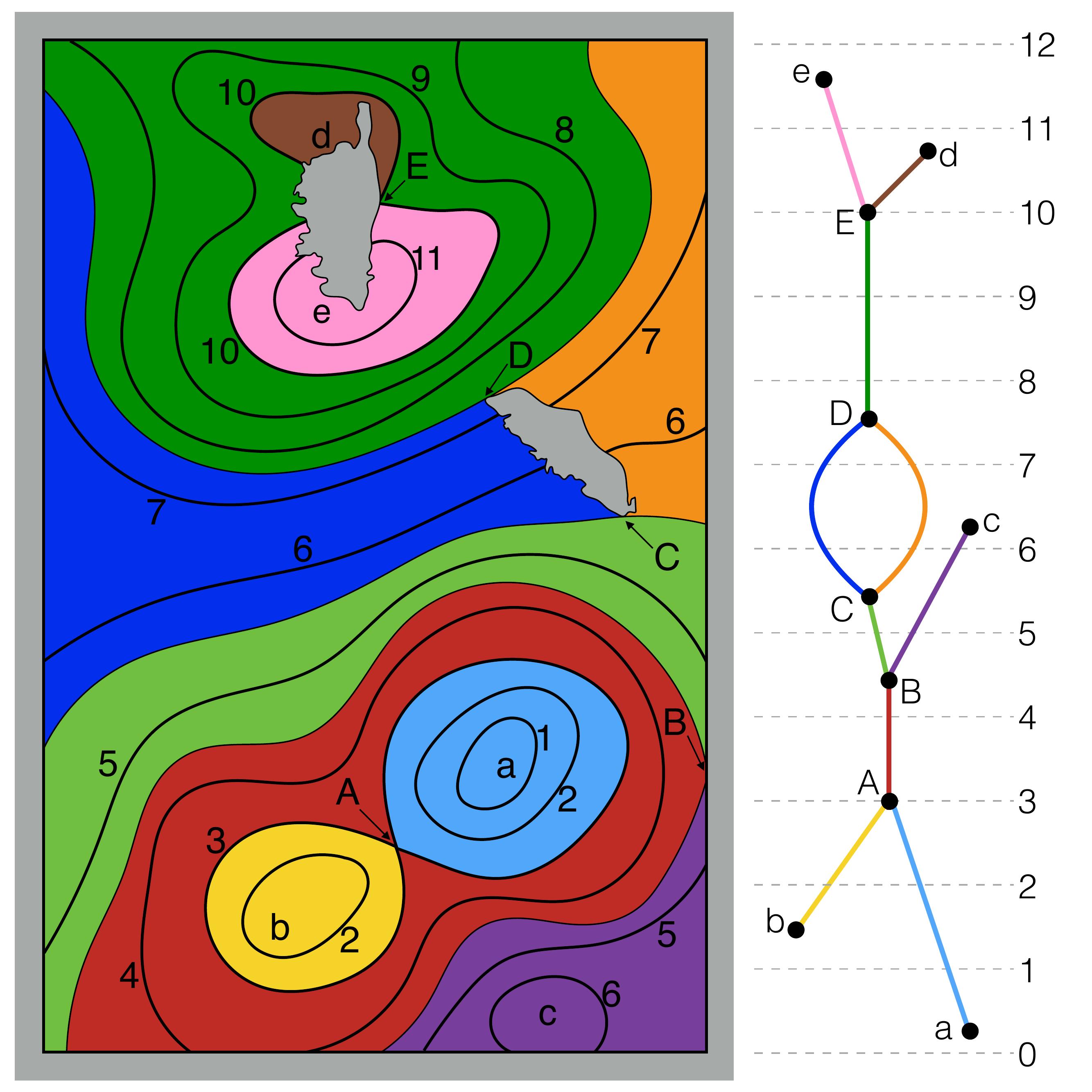}
  \caption{
  Contours (black, left) and Reeb graph (right) of $\protect\pn$, the pressure on a neutral surface, with grey islands. 
   Leaf nodes (small letters) indicate extrema of $\protect\pn$, while interior nodes (capital letters) indicate saddles of $\protect\pn$. 
   Each arc in the Reeb graph has an associated region in physical space, shown in matching colour. 
Islands (such as the bottom one, but not the top one) can create cycles in the Reeb graph.
For convenience, the nodes' vertical position is the pressure of their associated critical point; their horizontal position is arbitrary. 
}
\label{fig:schematic}
\end{figure}

To get started, first consider a region in $\M$ where $\nabla \pn \neq \bm{0}$, and where there is precisely one contour of $\pn$ for any given pressure value.
For example, consider just the light blue region in Fig.~\ref{fig:schematic}, surrounding point \node{a} and with $\pn < 3$. 
Since $\nabla \pn$ is orthogonal to a contour of constant $\pn$, and similarly for $\nabla \rn$ and $\rn$, \eqref{eq:def_ntp_svp} implies that the contour of $\pn$ through any point is parallel to the contour of $\rn$ through the same point.
Since this is true at all points, a contour of constant $\pn$ must be exactly a contour of constant $\rn$: the two contours are the same set of points.
Specifying a value of $\pn$ specifies a unique (by assumption) contour of $\pn$, which is identical to a contour of $\rn$, upon which $\rn$ is constant. 
Thus, there is a single-valued functional relationship between $\rn$ and $\pn$: \begin{equation}
\label{eq:svfn}
\rn = \rfn(\pn)
\end{equation}
for some function $\rfn$.
 
Now, the gradient of \eqref{eq:svfn} yields
\begin{equation}
\label{eq:gradsvfn}
\nabla \rn = \dydx{\rfn}{p}(\pn) \, \nabla \pn.
\end{equation}
Together, \eqref{eq:def_ntp_svp} and \eqref{eq:gradsvfn} require 
\begin{equation}
\label{eq:dsvfndp}
\dydx{\rfn}{p}(\pn) = \rpn.
\end{equation}
Not only does this give us $\mathrm{d} \rfn / \mathrm{d} p$, it says that $\rpn$ is also a function of $\pn$. Specifically,
\begin{equation}
\label{eq:svpfn}
\rpn = \rpfn (\pn),
\end{equation}
where $\rpfn(\parg) = (\mathrm{d} \rfn / \mathrm{d} p)(\parg)$. 
(The type-face $\parg$ distinguishes the function argument from the 3D scalar field $p$; this will be more necessary later.)
This is also evident by cross-differentiating \eqref{eq:def_ntp_svp}, to get
\begin{equation}
\label{eq:J_svpfn_p}
0 = \khat \cdot \nabla \rpn \times \nabla \pn.
\end{equation}
Thus $\nabla \rpn$ and $\nabla \pn$ are parallel, and the preceding logic applies.\footnote{For the notational convenience of \eqref{eq:J_svpfn_p}, $\nabla \rpn$ and $\nabla \pn$ are temporarily embedded in 3D space, with zero component in the $\khat$ direction. A more rigourous notation is $0 = J(\rpn \, , \, \pn)$, where $J$ is the Jacobian. 
} 
The relation between $\rfn$ and $\rpfn$ can equivalently be expressed by integrating $\rpfn$ to obtain
\begin{equation}
\label{eq:svpfn_int}
\rfn(\parg) = \rho_c + \int^{\parg}_{p_c} \rpfn(p') \, \mathrm{d} p',
\end{equation}
for some constant pressure $p_c$ and constant density $\rho_c = \rfn(p_c)$. 

\subsection{Multivalued functional relations}
In general $\nabla \pn = \bm{0}$ in some places, and in general there are multiple disjoint contours for a given value of $\pn$. 
Recognizing this, the logic that led from \eqref{eq:def_ntp_svp} to \eqref{eq:svfn} now reveals that $\rn$ can take different values on each of these disjoint contours, so $\rfn$ in \eqref{eq:svfn} is actually a \emph{multivalued} function of pressure. 

For example, now consider the region in Fig.~\ref{fig:schematic} with $\pn < 7$. 
The value of $\rn$ must be constant on the $\pn = 2$ contour surrounding point \node{a}, but can be different than the constant value of $\rn$ on the $\pn = 2$ contour surrounding point \node{b}. Similarly, there are three disjoint contours of the level set $\pn^{-1}(6)$, upon each of which $\rn$ may take a different value. 

At this point, the reader who wishes to see the oceanographic relevance of this multivalued relationship may jump to Section~\ref{sec:OCCA1}, taking in Fig.~\ref{fig:illustrate}a,b.

Our task is to determine geographic regions such that any level set of $\pn$ has no more than one connected component (contour) in each region, and thus only one value of $\rn$. 
Then, there are \emph{single-valued} functions that satisfy \eqref{eq:svfn} within each region. These regions, their meeting points, and the ways they nest into a global structure are encoded by the Reeb graph.  

\subsection{The Reeb graph}
The \citet{reeb1946} graph captures the essential topological information about connectedness of level sets of a real-valued function on a topological space.
For our purposes, the function is $\pn$ and the topological space is $\M$. 
Each contour of $\pn$ is contracted to a single point in the Reeb graph of $\pn$.\footnote{
More formally, the \textdef{Reeb graph} of $\pn : \M \rightarrow \mathbb{R}$ is the quotient space $\M / \!\! \sim$, where the equivalence relation $\sim$ is such that $\bm{x} \sim \bm{y}$ if $\bm{y} \in \M$ is an element of the contour of $\pn$ through $\bm{x} \in \M$.
}

The Reeb graph of $\pn$ in the preceding example is shown on the right of Fig.~\ref{fig:schematic}. 
In the preceding example, $\pn^{-1}(2)$ has two disjoint components (contours), so the Reeb graph of $\pn$ has two corresponding points at $\pn = 2$. 
Now consider the $\pn=2$ contour in the light blue region surrounding point \node{a}. Moving along a path in physical space from a point on this contour to lower pressure traces out the light blue curve in the Reeb graph, until this ends when the path reaches the pressure minima in physical space located at \node{a}.
Or consider the $\pn$ contours as $\pn$ is swept upwards towards $\pn=3$: the two $\pn$ contours approach each other, finally merging into a single contour at $\pn=3$. The $\pn=3$ contour contains the saddle point \node{A}, which joins three different curves in the Reeb graph. 

These points in the Reeb graph are most usefully expressed by a collection of $N$ nodes and $A$ arcs---a graph. 
Each node $n$ represents a geographic location $\bm{x}_n \in \M$ that is a critical point of $\pn$. Leaf nodes (nodes of degree one) represent local maxima or minima of $\pn$, and internal nodes (nodes with degree two or more) represent saddles of $\pn$. 
Denote the critical value of node $n$ by $p_n = \pn(\bm{x}_n)$.
An arc $a$ is incident upon two nodes, denoted $\ell_a$ and $h_a$ (think ``low'' and ``high''), having $p_{\ell_a} < p_{h_a}$, if 
 there is a path $\mathcal{P}_a$ from $\bm{x}_{\ell_a}$ to $\bm{x}_{h_a}$
that is strictly increasing in $\pn$ [\ie $s < t$ implies $\pn(\mathcal{P}_a(s)) < \pn(\mathcal{P}_a(t))$] and $\mathcal{P}_a(t)$ is not on the contour of any critical point of $\pn$ for all $t \in (0,1)$. 
Nothing topologically important happens on these paths between pairs of critical points. The region swept out by the contours intersecting such a path is the \textdef{(associated) region} $\reg_a \subseteq \M$ to arc $a$. Mathematically, $\reg_a = \cup_{t \in [0,1]} \ \pn^{-1} ( \mathcal{P}_a(t) )$. In Fig.~\ref{fig:schematic}, these are the coloured regions in physical space corresponding to the coloured arcs in the Reeb graph. 
Now consider points \node{C} and \node{D} in Fig.~\ref{fig:schematic}. There are infinitely many paths through physical space between these points, but all paths that go left of the island have the same associated region. Similarly all paths that go right have the same associated region. Thus the Reeb graph has two arcs between nodes \node{C} and \node{D}, which together form a cycle. 
In general, the Reeb graph has as many arcs between two given nodes as there are paths ($\mathcal{P}_a$ as above) with distinct associated regions. Such paths are non-homotopic, meaning they cannot be continuously deformed into one another while remaining in $\M$. 

As an aside, do not look too closely at the complex boundaries around the islands in Fig.~\ref{fig:schematic}. There ought to be many extrema of $\pn$ lurking around these boundaries, but for illustrative purposes these are ignored. Complex boundaries do add considerable complexity to the Reeb graph for real oceanic data.

The multivalued functions $\rfn$ and $\rpfn$ become single-valued when the domain is restricted to an associated region. 
For each arc $a$ in the Reeb graph, there is a single-valued function $\rfn_a : [p_{\ell_a},  p_{h_a}] \rightarrow \mathbb{R}$
such that 
\begin{equation}
\label{eq:svfn_a}
\rn(\bm{x}) = \rfn_a \big(\pn(\bm{x})\big) \quad \forall \bm{x} \in \reg_a.
\end{equation}
The functions $\rfn_a$ are called \textdef{branches} of the multivalued function $\rfn$.
In fact, for a perfectly neutral surface, $\rfn_a$ is \emph{defined} by the data $\{ (\pn(\bm{x}), \rn(\bm{x})) : \bm{x} \in \reg_a \}$.
The branches of $\rpfn$ are defined similarly, but extra care is needed at saddle points, discussed next.

To see the Reeb graph and its associated regions on oceanographic data, the reader may jump to Section~\ref{sec:OCCA2}, taking in Fig.~\ref{fig:illustrate}c,d.

\subsection{Pressure saddles}
\label{sec:pressure_saddles}

How do the branches of the multivalued functions $\rfn$ relate at the saddle pressures, and similarly for $\rpfn$?

For $\rfn$, the logic leading to \eqref{eq:svfn} applies perfectly well at saddle points. Contours of $\pn$ are contours of $\rn$, so $\rn$ is constant along contours of $\pn$, including those through saddle points. Thus, the branches of $\rfn$ must match continuously at saddle points. 
That is, for every internal node $s$, $\rn(\bm{x}_s) = \rfn_a\big(\pn(\bm{x}_s) \big)$ for all arcs $a$ incident upon node $s$.
Actually, critical points are contained in the sets $\reg_a$, so this condition is already covered by \eqref{eq:svfn_a}. 

For $\rpfn$, however, the logic leading to \eqref{eq:dsvfndp} does not apply at saddle points. Combining \eqref{eq:def_ntp_svp} and \eqref{eq:svfn} actually gives $(\mathrm{d} \rfn / \mathrm{d} p)(\pn) \, \nabla \pn = \rpn \nabla \pn$, trivially satisfied where $\nabla \pn = \bm{0}$.
Is it possible that the branches of $\rpfn$ do not match continuously at the saddle pressures?
The logic leading to \eqref{eq:dsvfndp} does apply everywhere on the $\pn$ contour through the saddle, except at the saddle itself. If we remove the saddle point from this contour, then it has multiple connected components, upon each of which $\rpn$ is constant. To rephrase the previous question, could $\rpn$ take different values on these different components?
If the answer were yes, then $\rpn$ would jump discontinuously at this point along a trajectory of constant pressure, which implies a discontinuity of salinity or temperature at this point. Assuming the ocean hydrography is continuous, the answer is no: the $\rpfn$ branches must match continuously at the saddle points.

However, rather than a saddle point, now imagine a saddle \emph{region} within which $\pn = p_s$, a constant.
As $\nabla \pn = \bm{0}$ here, \eqref{eq:def_ntp_svp} requires $\nabla \rn = \bm{0}$ too, so $\rn$ is constant within this region, and the functional relationship $\rfn$ holds. 
However, \eqref{eq:J_svpfn_p} is trivially satisfied here, even for non-zero $\nabla \rpn$, so $\rpn$ may be non-constant in this region. 
This entire ``flat'' region is a node in the Reeb graph. The branches of $\rfn$  for the arcs incident upon this node must agree at $p_s$, but those branches of $\rpfn$ may be undefined at $p_s$, and will disagree in the limit as the pressure approaches $p_s$. 
So, the multivalued function $\rfn$ must be continuous, but $\rpfn$ may have jump discontinuities at saddle regions.

Ultimately, $\rn$ and $\rpn$ are determined by $\sn$, $\tn$, and $\pn$, so it is helpful to discuss the structure of salinity and potential temperature on neutral surfaces.
Expanding $\nabla \rpn$ in \eqref{eq:J_svpfn_p} by the chain rule, akin to \eqref{eq:gradsv}, and using \eqref{eq:def_ntp_st2} to re-write $\nabla \sn = - \rtn \rsn^{-1} \nabla \tn$, we find
\begin{equation}
\label{eq:J_t_p}
0 = \left( \utilde{\rho_{p \theta}} - \utilde{\rho_{p S}} \, \rtn \, \rsn^{-1} \right) \khat \cdot \nabla \tn \times \nabla \pn.
\end{equation}
The term in parentheses is non-zero, and \eqref{eq:J_t_p} is of the form \eqref{eq:J_svpfn_p}, not the form \eqref{eq:def_ntp_svp}.
So, there is a multivalued function $\hat{\theta}$ for which $\tn = \hat{\theta}(\pn)$ with the same behaviour as $\rpfn$, including the possible jump discontinuities. 
Of course, one can equally well consider salinity, and find a multivalued function $\hat{S}$ for which $\sn = \hat{S}(\pn)$, again with the same behaviour as $\rpfn$.  
This is how $\rpn$ can vary inside a flat pressure region: in such a region, $\sn$ and $\tn$ can vary in compensatory ways to maintain $\rn$ constant, but cannot simultaneously maintain $\rpn$ constant. 
Rest assured that, while $\rpfn$, $\hat{\theta}$, and $\hat{S}$ may be discontinuous multivalued functions of pressure, the fields $\rpn$, $\tn$, and $\sn$ are themselves continuous in space. 

As a physical field, not a mathematical construction, we might expect there are no extended regions where $\pn$ is truly constant. If so, the branches of $\rpfn$ meet continuously and are given by $\rpfn_a : [p_{\ell_a},  p_{h_a}] \rightarrow \mathbb{R}$ such that 
\begin{equation}
\label{eq:svpfn_a}
\rpn(\bm{x}) = \rpfn_a \big(\pn(\bm{x})\big) \quad \forall \bm{x} \in \reg_a,
\end{equation}
analogous to \eqref{eq:svfn_a}.
Where a saddle $\bm{x}_s$ has constant pressure over an extended region, for each arc $a$ incident upon node $s$, simply exclude $p_s$ from the domain of $\rpfn_a$ and restrict $\bm{x}$ to the interior of $\reg_a$. 

A final, technical point is that the strict definition of the Reeb graph is rooted in Morse theory, which requires all critical points (of $\pn$) to be non-degenerate. This and other properties of Morse functions ensure all nodes of the Reeb graph are degree one or three. 
Of course, a region of constant $\pn$ is full of degenerate critical points.  Moreover, piecewise-linear functions constructed from numerical data often fail to be Morse. Nonetheless the Reeb graph still exists and can be computed, though it may have nodes with degree 2 or 4 or more \citep{cole-mclaughlin.edelsbrunner.ea2003, doraiswamy.natarajan2013}.

\subsection{Islands and holes and cycles}
\label{sec:islands}
Islands and other holes in the neutral surface can impose an additional constraint on $\rpfn$.
Consider a neutral trajectory, a path $\mathcal{P}$ in the neutral surface. The change of \emph{in-situ} density from (a reference point) $\bm{x}_\rv \in \M$ to (any point) $\bm{x} \in \M$ is
\begin{equation}
\label{eq:intsvpath}
\rn(\bm{x}) - \rn(\bm{x}_\rv) = \int_\mathcal{P} \nabla \rn \cdot \mathrm{d} \bm{r}
= \int_\mathcal{P} \rpn \nabla \pn \cdot \mathrm{d} \bm{r}.
\end{equation}
using \eqref{eq:def_ntp_svp} for the second equality. 
Having assumed the neutral surface is well-defined, these integrals are path-independent.

The path $\mathcal{P}$ in \eqref{eq:intsvpath}
corresponds to a walk 
 in the Reeb graph of $\pn$, alternately passing along arcs and through nodes in the order $a_0$, $n_1$, $a_1$, ..., $n_\mathcal{J}$, $a_\mathcal{J}$. 
Specifically, $\bm{x}_\rv \in \reg_{a_0}$ and $\bm{x} \in \reg_{a_\mathcal{J}}$. 
From \eqref{eq:svfn_a}, we also have $\rn(\bm{x}_\rv) = \rfn_{a_0}(p_\rv)$ and $\rn(\bm{x}) = \rfn_{a_\mathcal{J}}(p)$,  where $p_\rv = \pn(\bm{x}_\rv)$ and $\parg = \pn(\bm{x})$. 
Thus the path integral in \eqref{eq:intsvpath} becomes a ``graph integral'',

\begin{alignat}{3}
\label{eq:svpfn_int2}
\rfn_{a_\mathcal{J}}(\parg) 
= \rfn_{a_0}(p_\rv) \quad
& + & & \int_{p_\rv}^{p_{n_1}} && \rpfn_{a_0}(p') \, \mathrm{d} p' \nonumber \\
& + & \quad \sum_{j=1}^{\mathcal{J}-1} &\int_{p_{n_j}}^{p_{n_{j+1}}} && \rpfn_{a_j}(p') \, \mathrm{d} p' \nonumber \\
& + &  &\int_{p_{n_{\mathcal{J}}}}^{\parg} && \rpfn_{a_{\mathcal{J}}}(p') \, \mathrm{d} p',
\end{alignat}
having also used \eqref{eq:svpfn_a}. This generalizes \eqref{eq:svpfn_int} when the functions $\rfn$ and $\rpfn$ are multivalued.

The only way path-dependence can affect \eqref{eq:svpfn_int2} is via cycles in the Reeb graph. 
To see this, suppose the Reeb graph has no cycles (a tree), and consider any two nodes. There is a unique walk between these nodes having no repeated nodes, called a straight walk. 
 The result of \eqref{eq:svpfn_int2} for any walk between these nodes is identical to that for the straight walk, so \eqref{eq:intsvpath} is path-independent.

If there are holes in $\M$ (such as created by islands), there may be cycles in the Reeb graph. In this case, there may be multiple straight walks between two nodes, corresponding to the path $\mathcal{P}$ navigating one way or another around a hole.
Consider a cycle whose walk is $n_1$, $a_1$, $n_2$, $...$, $n_{\mathcal{J}-1}$, $a_{\mathcal{J}-1}$, $n_\mathcal{J} = n_1$.
Without loss of generality, suppose $\bm{x}_\rv = \bm{x} = \bm{x}_{n_1}$.
Then \eqref{eq:svpfn_int2} becomes
\begin{equation}
\label{eq:island}
0 = \sum_{j=1}^{\mathcal{J}-1} \int_{p_{n_j}}^{p_{n_{j+1}}} \rpfn_{a_j}(p') \, \mathrm{d} p' .
\end{equation}
This constraint must be satisfied by $\rpfn$ for each cycle in the Reeb graph of $\pn$.

It may come as a surprise that the existence of holes in $\M$ does not immediately guarantee that the Reeb graph of $\pn$ will have cycles. A cycle can only occur if there is a $\pn$ contour that intersects the hole at precisely one end; the contour cannot close on itself, so its other end must intersect a different hole. 
If only one hole is contained within a $\pn$ contour (such as the top island in Fig.~\ref{fig:schematic}) or between two $\pn$ contours, that hole does not produce a cycle:
any contour intersecting the hole cannot cross the bounding contour(s), and so must intersect the hole twice.
As a corollary to this, there are no cycles when $\M$ has a single hole. 
Note that a bounded ocean containing a single island, as in Fig.~\ref{fig:island_helix}, is topologically the same as an aqua-planet with two islands.

It is conceivable that the pitch of neutral helices in the open ocean is small ($H \approx 0$ globally), yet around islands or other holes this pitch may be large. 
This seems plausible if interior ocean dynamics naturally tend to destroy (bring to zero) neutral helicity, as \citet{mcdougall.jackett2007} tentatively suggested.  
A preliminary analysis of this possibility is given in \ref{sec:island_pitch}. 

\subsection{Summary}

When $H = 0$ everywhere, neutral surfaces are well-defined surfaces. 
On a well-defined neutral surface, the \emph{in-situ} density, $\rn$, is a multivalued function of the pressure, $\pn$. The neutral surface can be covered by regions within each of which $\rn$ is a single-valued function of $\pn$, called a branch of the multivalued function. 
Each of these regions, and so too each of the branches, is associated with an arc of the Reeb graph of $\pn$.
The matching conditions of these branches are determined by the structure of the Reeb graph. 
Each internal node of the graph is associated with a saddle of $\pn$.
All branches associated with arcs incident to a common node must match continuously at the pressure associated with that node.

Moreover, the partial derivative of \emph{in-situ} density with respect to pressure on a well-defined neutral surface, $\rpn$, is also a multivalued function of $\pn$. It is single-valued within the same regions as above and these branches must integrate to zero around every cycle in the Reeb graph.

Following an oceanographic example in Section~\ref{sec:illustrative}, this theory will be used in Section~\ref{sec:topobaric} to develop unique-depth, approximately neutral surfaces in the real ocean with $H \neq 0$. 

\section{Illustrative example}
\label{sec:illustrative}

\subsection{Pressure and \emph{in-situ} density on an approximately neutral surface}
\label{sec:OCCA1}

The main computations and tests presented will use high resolution data (spatially and temporally), but for illustrative purposes smoother fields are desired: the OCCA 2004---2006 climatology \citep{forget2010} provides these.
The potential density referenced to \SI{1000}{dbar} is (crudely) calculated using the climatological salinity and potential temperature, then the isopycnal surface intersecting ($\xref$, $\yref$, \SI{1000}{dbar}) is found. Then, the pressure on this isopycnal, $\pn$, is slightly adjusted to globally minimize the error from neutrality, resulting in an $\omega$-surface \citep{klocker.mcdougall.ea2009}. This is the ``illustrative surface''. 
This calculation is crude, essentially treating climatological data as if it were instantaneous, but suffices for illustration. It also ignores the fact that OCCA is a Boussinesq model (the implications of which will be discussed in Section~\ref{sec:data}). 

\begin{figure*}[p]
\centering
\includegraphics[width=1\textwidth]{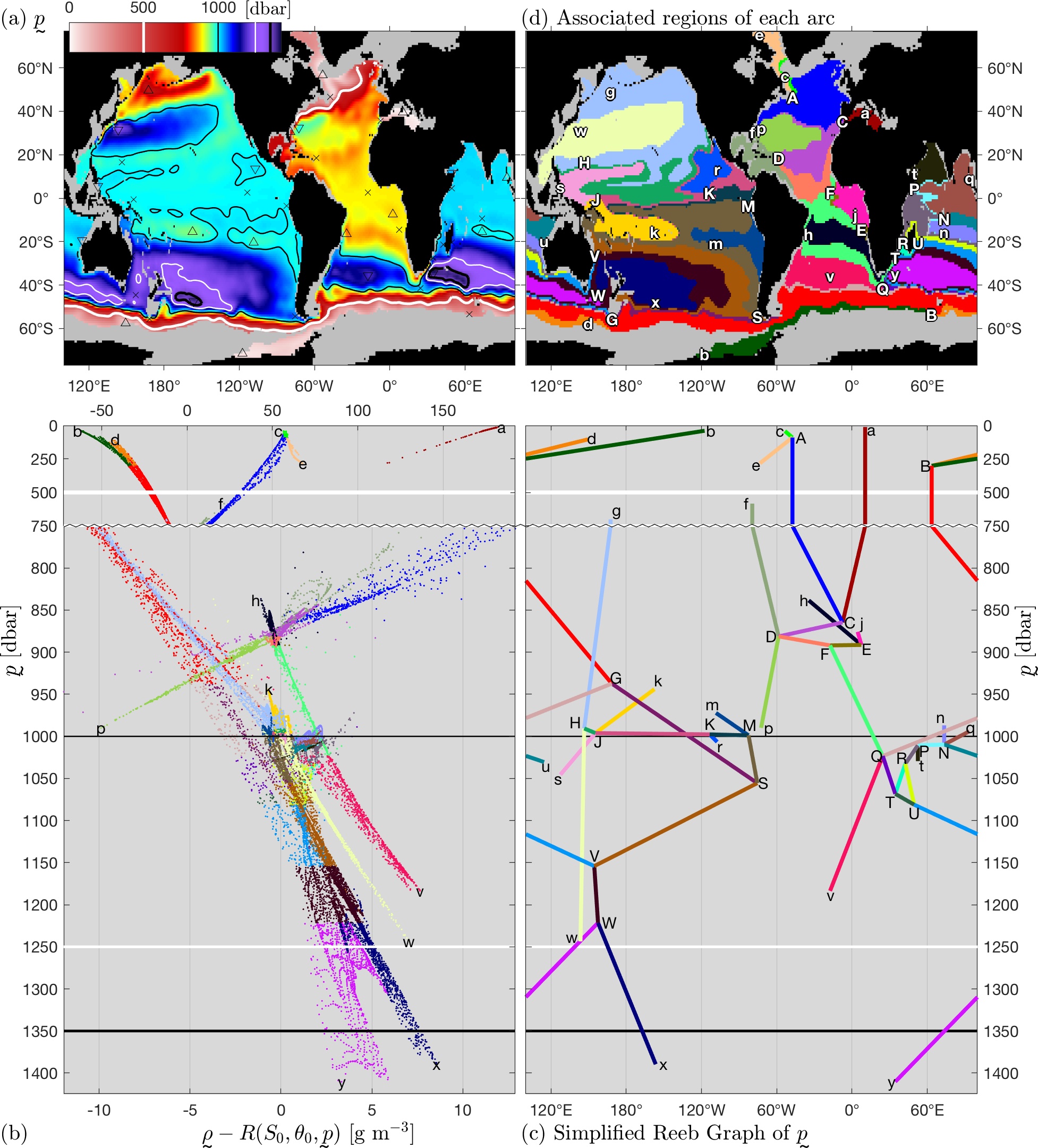}
\caption{The illustrative $\omega$-surface.
(a) The pressure $\protect\pn$ on the surface, also showing level sets of $\protect\pn$ at 500 (thick white), 1000 (thin black), 1250 (thin white), and \SI{1350}{dbar} (thick black). 
(b) Pressure $\protect\pn$ \vs \emph{in-situ} density $\protect\rn$ on the surface, with a reference profile $\eos(S_\rv, \theta_\rv, \protect\pn)$ subtracted from $\protect\rn$, purely for illustrative purposes.
Note a restricted abscissa is shown for pressure greater than \SI{750}{dbar}, where the ordinate spacing changes (serrated line).
(c) The simplified Reeb graph of $\protect\pn$, with nodes positioned at their corresponding critical point in pressure-longitude space, and labelled by letters: upper case for internal nodes (saddles) and lower case for leaf nodes (extrema).
(d) Associated regions of each arc of the simplified Reeb graph, with nodes indicated at their corresponding critical point in latitude-longitude space. 
Colour is coordinated between (b), (c), and (d). 
The level sets in (a) are indicated by horizontal lines in (b,c).
Grey regions in (a,d) indicate the surface is outcropped, incropped, or disconnected from the main ocean.
Certain nodes are also shown in (b).
In (a), symbols $\times$, $\bigtriangledown$, and $\bigtriangleup$ indicate those saddles, maxima, and minima (respectively) of $\protect\pn$ that correspond to nodes in the simplified Reeb graph.
}
\label{fig:illustrate}
\end{figure*}

The pressure on the illustrative surface, $\pn$, is mapped in Fig.~\ref{fig:illustrate}a. The (thick white) $\pn = \SI{500}{dbar}$ level set possesses two clearly disjoint contours in the Southern Ocean and North Atlantic.
If this surface were truly neutral, the \emph{in-situ} density would be constant along each of these pressure contours, yet possibly different between them. 
This phenomenon gives rise to the multivalued nature of $\rfn$ evident from the scatter plot of $\rn$ \vs $\pn$ shown in Fig.~\ref{fig:illustrate}b. (A reference profile $\eos(S_\rv, \theta_\rv, \pn)$ is subtracted from $\rn$ purely for illustrative purposes, as $\rn$ \vs $\pn$ looks essentially linear. The values $S_\rv$ and $\theta_\rv$ are taken as $\sn$ and $\tn$ at ($\xref$, $\yref$).)
Indeed, at $\pn = \SI{500}{dbar}$, $\rn$ is very nearly one of two values, each corresponding to one of the two disjoint contours of $\pn = \SI{500}{dbar}$.
Actually, there is some scatter of $\rn$ around these values. A potential density surface would show more scatter, whereas a truly neutral surface would show no scatter whatsoever---but the essential multivalued nature would remain.

\subsection{The Reeb graph and its associated regions}
\label{sec:OCCA2}

The Reeb graph of $\pn$ for the illustrative surface contains 1,370 arcs---too many to be particularly informative. Most of these represent very small regions in physical space, often hugging coasts where local extrema of $\pn$ are common. Even with climatological fields, the Reeb graph requires simplification. 
This is a complicated task, and used only for this illustration, so only a brief description is given here; see \citet{stanley2018thesis} for further details. 
In brief, small holes are filled in with extremely large values (so contours go around them), then the Reeb graph is calculated, then those filled holes removed from the associated regions. 
The simplification method that \citet{carr.snoeyink.ea2010} used on trees is then used on this graph with cycles. It iteratively removes the least important leaf node until only a specified number of arcs remain; it never destroys cycles and never produces a node with only arcs leading to lower pressures, or only arcs leading to higher pressures. 

The Reeb graph of $\pn$ on the illustrative surface is computed and simplified down to 43 arcs. 
The graph itself is drawn in Fig.~\ref{fig:illustrate}c with nodes positioned according to the pressure (ordinate) and longitude (abscissa) of their associated critical points.\footnote{This graph drawing conveys additional useful information at the cost of some arcs overlapping. Deciding the best placement of nodes and arcs of a Reeb graph is non-trivial \citep{heine.schneider.ea2011}.}
The regions $\reg_a$ associated with each arc $a$ of this simplified Reeb graph are mapped in Fig.~\ref{fig:illustrate}d. 
Three arcs cross the (thick white) \SI{500}{dbar} level. Each of these supports a distinct \emph{in-situ} density, as seen in Fig.~\ref{fig:illustrate}b.
(At OCCA's coarse-resolution, Mediterranean outflow on this surface jumps from \SI{280}{dbar} to \SI{850}{dbar} between neighbouring grid cells, so the \SI{500}{dbar} contour in Fig.~\ref{fig:illustrate}a is hardly visible near Gibraltar, though it exists.)
The associated regions for these arcs lie in
the Southern Ocean (red), 
the North Atlantic (blue), 
and the Mediterranean (maroon).
The point on the Reeb graph on the blue arc at \SI{500}{dbar} represents the entire, but single, \SI{500}{dbar} contour of $\pn$ in North Atlantic region. 
 Following this arc upwards corresponds to a path, on the surface, in which the pressure decreases monotonically. It can be followed until \SI{90}{dbar}, where node \node{A} is reached, corresponding to a saddle of $\pn$ at the Grand Banks of Newfoundland (indicated by \node{A} in Fig.~\ref{fig:illustrate}d). All such pressure-monotonic paths in the North Atlantic between \SI{90}{dbar} (node \node{A}) and \SI{865}{dbar} (node \node{C}) are equivalent in the Reeb graph. 
As pressure increases along these paths, the \emph{in-situ} density increases identically between all such paths (were this surface truly neutral). At the saddle \node{A}, there are two options: the path can be continued monotonically to shallower pressure, ultimately reaching the local minima of $\pn$ in the North Atlantic (node \node{c}); or, the path can descend to higher pressure, ultimately reaching the local maxima of $\pn$ in Baffin Bay (node \node{e}). 

Similarly, the point on the Reeb graph on the dark blue arc at \SI{1350}{dbar} represents the entire, but single, \SI{1350}{dbar} contour in the South Pacific. 
As one moves along this arc to node \node{x}, the contour contracts to a point, the maxima of $\pn$ in the South Pacific. Or, moving to shallower pressures, the contour grows, until it intersects another contour of the same level set from the South Indian, at the saddle node \node{W}. 
Note, at \SI{1250}{dbar} in the South Pacific, the \emph{in-situ} density seems to branch into two distinct values. This is because a small closed $\pn$ contour, west of New Zealand, has been added to the dark blue region by the simplification process (compare Fig.~\ref{fig:illustrate}a and d).  Much of the scatter in Fig.~\ref{fig:illustrate}b is actually due to such merging of regions by the simplification process, rather than poor neutrality of the $\omega$-surface.

Now consider an upward-monotonic path from the South Pacific $\pn$ maxima (node \node{x}) to the Amundsen Sea (node \node{b}). The most direct route is equivalent, in the Reeb graph, to a journey around the South Pacific visiting nodes \node{W}, \node{V}, \node{S}, \node{G}, \node{B}, then \node{b}.
Alternatively, one can take a South Indian route, through nodes \node{W}, \node{V}, \node{U}, \node{T}, \node{Q}, \node{G}, \node{B}, then \node{b}. These are two ways around New Zealand. As New Zealand (united, at this depth) represents a hole in $\M$, it creates a cycle in the Reeb graph: \node{V}, \node{S}, \node{G}, \node{Q}, \node{T}, \node{U}, \node{V}.
Madagascar creates a second cycle in the Reeb graph, \node{U}, \node{T}, \node{R}, \node{U}.
A third, and final, cycle is the loop around both New Zealand and Madagascar.
There are many more holes in $\M$, but only these cycles.
Antarctica does not produce a cycle in the Reeb graph, because there exists a contour of $\pn$ encircling Antarctica, as discussed in Section~\ref{sec:islands}.
(One can, equivalently, think of the non-Antarctic continents as islands in an ocean bounded by Antarctica.)
The simplification procedure pre-emptively removed other, smaller holes before calculating the Reeb graph.
No such simplification will be used when computing topobaric surfaces, discussed next.

\section{Topobaric surfaces}
\label{sec:topobaric}

Studying the topology of a neutral surface in an ocean with zero neutral helicity (Section~\ref{sec:topology_neutral_surfaces}) revealed the existence of a multivalued function $\rfn$ whose branches satisfy \eqref{eq:svfn_a}. 
The real ocean has neutral helicity that is non-zero yet small, so surfaces can only be approximately neutral. On such surfaces, $\nabla \rn - \rpn \nabla \pn$ is non-zero, but small enough that the multivalued function $\rfn$ still usefully describes the approximately neutral surface, albeit a small error must be added to \eqref{eq:svfn_a}; this error is the scatter in Fig.~\ref{fig:illustrate}b. 

Topobaric surfaces turn this around, by forcing \eqref{eq:svfn_a} to be exact for a given $\rfn$. To obtain nearly neutral surfaces, $\rfn$ must be chosen carefully. Clearly, choosing a constant $\rho_c = \rfn(p)$ is a bad choice. 
The theory for how to choose $\rfn$ is given next (Section~\ref{sec:topobaric_theory}), followed by an algorithm to construct topobaric surfaces (Section~\ref{sec:topobaric_methods}). This method is applied to data (described in Section~\ref{sec:data}) to numerically compute topobaric surfaces and compare them with other approximately neutral surfaces (Section~\ref{sec:topobaric_results}).

\subsection{Theory}
\label{sec:topobaric_theory}

Because the neutral helicity is non-zero, and because we wish to make unique-depth surfaces in the presence of multiple islands and other holes in $\M$, we cannot construct perfectly neutral surfaces.
Starting in the simple setting of Section~\ref{sec:singlevalued_functional_relations} where $\rfn$ and $\rpfn$ are single-valued, this means that a unique-depth surface cannot simultaneously satisfy the $\rfn$ relation \eqref{eq:svfn} and the $\rzfn$ relation \eqref{eq:svpfn} exactly. 
We could choose $\rfn$ upfront then force \eqref{eq:svfn} to hold exactly and not worry about \eqref{eq:svpfn}. But, for neutrality, what really matters is not the values of $\rfn$ but its derivative, as  \eqref{eq:dsvfndp} shows.
So, we choose to satisfy \eqref{eq:svfn} exactly and approximate \eqref{eq:svpfn} with an empirically fit $\rpfn$, which is integrated to determine $\rfn$ according to \eqref{eq:svpfn_int}. 
To see that this maximizes neutrality, combine \eqref{eq:svfn} and \eqref{eq:svpfn_int} and take the gradient via the Leibniz integral rule to find $\nabla \rn = \rpfn(\pn) \nabla \pn$.
The approximate version of \eqref{eq:svpfn} then yields $\rpfn(\pn) \nabla \pn \approx \rpn \nabla \pn$. 
Combining these gives an approximate version of the neutrality condition \eqref{eq:def_ntp_svp}. The better $\rzfn(\pn)$ approximates $\rpn$, the closer the resulting surface will be to neutral.\footnote{Another possibility would be to make \eqref{eq:svpfn} exact and obtain $\rpfn$ by differentiating an empirically fit $\rfn$ that approximates \eqref{eq:svfn}. But, as above when choosing $\rfn$ upfront, there is no reason to believe the surface will be nearly neutral. The condition \eqref{eq:J_svpfn_p} would be guaranteed, but this does not imply neutrality \eqref{eq:def_ntp_svp}.}

Translating this into the case of multivalued functions, 
a \textdef{topobaric surface}, denoted a $\tau$-surface, satisfies the $\rfn$ relation \eqref{eq:svfn_a} exactly, where $\rfn$ is obtained by integrating $\rpfn$ according to \eqref{eq:svpfn_int2} with a given a reference location $\bm{x}_\rv$, and $\rpfn$ satisfies the cycle constraint \eqref{eq:island}. 
To make topobaric surfaces as neutral as possible, $\rpfn$ is chosen to approximate \eqref{eq:svpfn_a} as best as possible.
The cycle constraint \eqref{eq:island} ensures topobaric surfaces are unique-depth surfaces.
Also, the exactness of \eqref{eq:svfn_a} ensures topobaric surfaces possess an exact geostrophic streamfunction \citep{stanley2019geostrf}. 

Topobaric surfaces allow $\rpfn$ to be discontinuous at all pressure saddles. 
There are three justifications for this. 
First, constant pressure regions could exist in the continuous $\pn$, but not be present in a discrete data representation of $\pn$.
Second, with non-zero neutral helicity \eqref{eq:J_svpfn_p} becomes $\epsilon = \khat \cdot \nabla \rpn \times \nabla \pn$ for some small scalar field $\epsilon$. Near saddle pressures $|\nabla \pn|$ is small, so $|\nabla \rpn|$ can be large while maintaining a small $\epsilon$, and thus $\rpn$ can change rapidly near saddle pressures. This differs from $\bm{\epsilon} = \nabla \rn - \rpn \nabla \pn$, which requires $|\nabla \rn|$ to be small near pressure saddles. 
Third, numerical tests show that requiring $\rpfn$ to match continuously at the pressure saddles produces surfaces that are less neutral.

This definition of topobaric surfaces is circular: $\pn$ must be known in order to calculate the Reeb graph of $\pn$, which is used in defining the multivalued functions $\rpfn$ and $\rfn$, the latter being an implicit definition for $\pn$ via \eqref{eq:svfn_a}. 

\subsection{Methods}
\label{sec:topobaric_methods}

To overcome the preceding circular definition, surfaces are built by an iterative algorithm that converges to a topobaric surface, as follows.
\begin{enumerate}
\item Begin with an approximately neutral surface, with pressure $\pn$, and a reference location $\bm{x}_\rv$.  
\item Compute the Reeb graph of $\pn$.
\item Empirically fit the branches $\rpfn_a$ using the data \linebreak $\{(\pn(\bm{x}), \rpn(\bm{x}) ) : \bm{x} \in \reg_a \}$ for each arc $a$ of the Reeb graph, subject to the cycle constraints \eqref{eq:island}.
\item Obtain $\rfn$ by integrating $\rpfn$ according to \eqref{eq:svpfn_int2}. 
\item Update $\pn$ with that satisfying \eqref{eq:svfn_a}, which is a root-finding problem for each water column.
\item Return to Step 2, unless a convergence test is passed.
\end{enumerate}
Surfaces constructed by a finite number of the above iterations are loosely referred to as topobaric surfaces. 
The remainder of this section describes each of these steps, in turn.

\subsubsection{Initial surface and reference location}

The initial surface can be any approximately neutral surface, chosen by the user. 
The resulting topobaric surface is somewhat dependent on this choice; starting from an $\omega$-surface yields slightly better results than starting from a potential density surface (not shown). 

Results shown in this paper all use topobaric surfaces initialized from potential density surfaces, to ensure the method succeeds when initialized from a surface that is not particularly neutral, globally. Also, one advantage of topobaric surfaces is their computational speed, which is defeated if one must first construct an $\omega$-surface. 

Next, the user selects a reference location $\bm{x}_\rv$. 
Then, by linear interpolation in the water column $\bm{x}_\rv$, record $p_\rv = \pn(\bm{x}_\rv)$, $S_\rv = \sn(\bm{x}_\rv)$, and $\theta_\rv = \tn(\bm{x}_\rv)$; also record $\rho_\rv = \eos(S_\rv, \theta_\rv, p_\rv) = \rn(\bm{x}_\rv)$.
The iteratively updated surface will maintain these properties at $\bm{x}_\rv$. (The initial potential density surface can have any reference pressure, but $p_\rv$ is a good choice.)

\subsubsection{Computing the Reeb graph}
Fast, robust, and general computation of Reeb graphs from discretised data has only been achieved recently.
\citet{carr.snoeyink.ea2003} developed a fast and general method to compute the contour tree (a Reeb graph with no cycles) in any dimension, based on sorting the data and sweeping through it twice. This is used by \citet{doraiswamy.natarajan2013} to compute the Reeb graph, by decomposing the space into a collection of loop-free regions. Their software, called ReCon, is written in Java. It is slightly modified to work with 64-bit floating point numbers, and to communicate directly with MATLAB.

ReCon requires its input to be a simplical mesh. This is a collection of vertices and, in 2D, a collection of triangles. With function values specified on the vertices, a piecewise-linear function may be constructed by linear interpolation. 
This differs from data on a rectilinear grid, as is common for oceanographic data. Rectilinear data may be bilinearly interpolated, but this non-linear interpolation can introduce new critical points, thereby changing the topology of level sets of $\pn$. 
How one constructs the simplical mesh from rectilinear data can matter \citep{carr.moller.ea2006}.
One method is to add a new vertex at the centre of each rectangle by four-way averaging, then split each rectangle into four triangles. This is how contours are typically computed from rectilinear data by the marching cubes algorithm. However, the associated regions for some arcs of the Reeb graph can contain only these extra vertices. To then fit $\rpn$ to $\pn$ would require calculating $\rpn$ on these extra vertices; this requires averaging $\sn$ and $\tn$ onto these extra vertices, but this is undesirable because $\rpn$ is a non-linear function of $\sn$, $\tn$, and $\pn$. We use a simple method, splitting each rectangle into two triangles; where a rectangle has one data point missing, only one triangle is produced. Using global oceanographic data, both methods produce Reeb graphs that appear similar. 

For simplicity, one connected component of the approximately neutral surface is handled at a time. A graph (not the Reeb graph) is constructed from the simplical mesh, having a node at every vertex of the mesh, and an arc between two nodes when their corresponding vertices share a face of the mesh. The connected component of this graph containing the point $\bm{x}_\rv$ is found \citep{tarjan1972}.\footnote{Faster, image manipulation methods are not sufficient because the simplical decomposition can remove the odd grid point, such as those with ground for 7 of 8 neighbours.}
 This guarantees that the Reeb graph is a connected graph.

Finally, all critical points must have unique values. Following standard practice \citep{doraiswamy.natarajan2013}, any duplicate values of $\pn$ are perturbed by a machine-precision amount until all vertices have unique values.

ReCon computes the Reeb graph in
$\mathcal{O}(v \log v + s n)$ time, where 
$v$ is the number of vertices,
$n$ is the number of triangles, and
$s$ is the number of saddles of the simplical mesh.

\subsubsection{Empirically fit  $\rpfn$, with cycle constraints}

The goal now is to use the data $\{(\pn(\bm{x}), \rpn(\bm{x})) : \bm{x} \in \reg_a \}$ to empirically fit a function $\rpfn_a$ that approximates \eqref{eq:svpfn_a}, for each arc $a$ of the Reeb graph of $\pn$, subject to the cycle constraint \eqref{eq:island} for each cycle. 

Rather than finding all cycles, we need only find a cycle basis, which is a minimal set of cycles out of which all other cycles can be produced by ``addition'': taking the arcs in one, but not both, of two cycles to produce a third cycle.\footnote{If the original two cycles are disjoint, the third is not a cycle but a more general object, an Eulerian subgraph---see \ref{sec:mathdefs_graphtheory}.} 
If \eqref{eq:island} holds for the first two cycles, it will hold for the third. A cycle basis is determined in two steps. 

First, find the arc $a_1$ having $\bm{x}_\rv \in \reg_{a_1}$, and let $m_1$ and $n_1$ be the nodes that $a_1$ is incident upon.  Perform a breadth-first search, starting with $m_1$ as the initially discovered node.
This iteratively discovers nodes adjacent to previously discovered nodes (the first step is rigged so that $n_1$ is discovered next). The result is a sequence of nodes
$ n_1, ..., n_{N-1} $ and another sequence of nodes $ m_1, ..., m_{N-1} $ such that $m_j$ was discovered before its adjacent node $n_j$. 
(The set $\{ m_1, n_1, .., n_{N-1} \}$ is all nodes in the graph.)
The set of arcs $\{ a_1, ..., a_{N-1} \}$, where $a_j$ is incident upon both $m_j$ and $n_j$, forms a spanning tree. 

Second, for each arc $a$ in the Reeb graph but not in the spanning tree, perform another breadth-first search in the spanning tree starting at a node upon which $a$ is incident, and stopping upon discovery of the other node upon which $a$ is incident. This finds the shortest walk in the spanning tree between the adjacent nodes of $a$. This, together with $a$, gives a cycle. All such cycles form the cycle basis.

Now, the branches $\rpfn_a$ can be empirically fit. 
What form should $\rpfn_a$ take?
The equation of state $\eos$ is usually expressed by a rational function of $S$, $\theta$, and $p$, so $\di_p \eos$ is also a rational function of $S$, $\theta$, and $p$. It might make sense to fit $\rpfn_a$ as a rational function of $p$,
 but a functional form with fewer degrees of freedom is preferable to avoid over-fitting the data. A form with two degrees of freedom will never be under-determined, because the chosen simplical mesh has at least two data points for each arc. Thus, we use the simple form
\begin{equation}
\label{eq:svpfn_form}
\rpfn_a(p) = K_a + L_a \, (p - p_{\ell_a}) + \di_p \eos(S_\rv, \theta_\rv, p),
\end{equation}
where the constants $K_a$ and $L_a$ are to be determined. The addition of $\di_p \eos(S_\rv, \theta_\rv, \pn)$ helps capture some of the non-linear behaviour of $\rpn$ with respect to $\pn$. 
For all arcs $a$ that are not in a cycle in the cycle basis, $K_a$ and $L_a$ are determined by ordinary least squares, fitting $K_a + L_a \, (\pn - p_{\ell_a})$ to $\rpn - \di_p \eos(S_\rv, \theta_\rv, \pn)$ within $\reg_a$.  
(That these may each be fit independently is a boon of allowing $\rpfn$ to meet discontinuously at the saddle pressures.)
The remaining branches are fit similarly, but as a single, coupled problem that also satisfies the cycle constraints \eqref{eq:island} for each cycle; this is done using MATLAB's \texttt{lsqlin} function. 
As the pressure difference between adjacent nodes is typically small, affine linear functions perform very well: in practice, the $\di_p \eos(S_\rv, \theta_\rv, p)$ term in \eqref{eq:svpfn_form} could be excluded with very little detriment.

\subsubsection{Obtain $\rfn$ by integrating $\rpfn$}
Having chosen the branches $\rpfn_a$ in \eqref{eq:svfn_form}, they are integrated according to \eqref{eq:svpfn_int2} to obtain the branches $\rfn_a$. 
In practice, each branch is first analytically integrated as in \eqref{eq:svpfn_int} to get 
\begin{equation}
\label{eq:svfn_form}
\rfn_a(p) = J_a + K_a (p - p_{\ell_a}) + \frac{L_a}{2} (p - p_{\ell_a})^2 + \eos(S_\rv, \theta_\rv, p).
\end{equation}
Each branch $a$ has a free constant of integration, $J_a$. 
One of these, $J_{a_1}$, is set by requiring $\rho_\rv = \rfn_{a_1}(p_\rv)$. The remainder are used to ensure the branches of $\rfn$ match continuously at the saddle pressures. This is done using the discovery order of the nodes from the previous step, as follows.
First, record $\rho_{m_1} = \rfn_{a_1}(p_{m_1})$. 
Then, for each $j = 2, ..., N-1$,
determine $J_{a_j}$ from $\rfn_{a_j}(p_{m_j}) = \rho_{m_j}$ and
record $\rho_{n_j} = \rfn_{a_j}(p_{n_j})$. 
Finally, for each arc $a$ in the cycle basis, 
determine $J_a$ from $\rfn_{a}(p_{\ell_a}) = \rho_{\ell_a}$.
The cycle constraint \eqref{eq:island} ensures this is identical to determining $J_a$ from $\rfn_{a}(p_{h_a}) = \rho_{h_a}$.

\subsubsection{Updating $\protect\pn$}

With all branches of $\rfn$ chosen, we must now update $\pn$ to satisfy \eqref{eq:svfn_a}. Specifically, for each arc $a$ and for each geographic position $\bm{x} \in \reg_a$, we set $\pn(\bm{x}) = p'$ where $p'$ 
solves
\begin{equation}
\label{eq:root_problem}
\eos\Big( \mathcal{S}(\bm{x}, p'), \ \vartheta(\bm{x}, p'), \ p' \Big) = \rfn_a(p'),
\end{equation}
where $\mathcal{S}$ and $\vartheta$ are versions of $S$ and $\theta$ with pressure as the vertical coordinate. Mathematically, $\mathcal{S}(\bm{x}, p') = S(\bm{x}, z')$ where $z'$ solves $p(\bm{x},z') = p'$, and similarly for $\vartheta$. 

Bisection is used to solve \eqref{eq:root_problem}. 
Since multiple solutions are possible, an initial guess is provided, based on $\pn$ from the previous iteration. A small interval around the initial guess is tested for a sign change at its limits. If a sign change is found, bisection proceeds inside this interval. Otherwise, the interval is (geometrically) expanded until a sign change is found and bisection can proceed, or the shallowest and deepest grid cells are reached. If the latter, no solution is found, meaning the updated surface has outcropped or incropped.\footnote{There is currently no capacity for ``wetting'', whereby subsequent iterations retest water columns that previously outcropped or incropped. This would require a way to define $\rfn$ for this water column. Perhaps $\rfn_a$ could be used when the water column is entirely surrounded by a single region $\reg_a$. At the boundary between regions, perhaps an average of these branches could be used.}
MATLAB's code generation is used to turn this into fast C executables (MEX).

\subsubsection{Iteration}
Recall that $\rpfn$ is allowed to be discontinuous only at the saddle pressures. The iterative method is required because solving for new pressures (step 5) causes the saddle points to change, and thus after an iteration, $\rpfn$ will be discontinuous \emph{inside} some regions. This causes large errors along the pressure contours that were formerly through pressure saddles. As the whole algorithm iterates, $\pn$ converges, and discontinuities of $\rpfn$ occur only at the pressure saddles. 

The stopping criterion may be chosen by the user. The default is to stop when the root-mean-square change of $\pn$ is less than $10^{-3}$ dbar. Provided this value is sufficiently small, the results are not sensitive to the choice of this stopping value.

\subsection{Data \& Boussinesq models}
\label{sec:data}
With these methods, topobaric surfaces are constructed and tested using ECCO2 \citep{menemenlis.hill.ea2005} data, having $0.25^\circ \times 0.25^\circ$ horizontal resolution, on 22--24 December 2002. A single archived time-step is chosen to make the task as hard as possible: the Reeb graph of a smoother climatology is considerably simpler. 
In truth, neutral helices possess a temporal as well as spatial dimension \citep{klocker.mcdougall2010}, which we are not considering here.

Boussinesq models, such as ECCO2, calculate the \emph{in-situ} density 
from a Boussinesq equation of state $\eosB$ that uses depth $z$ (negative and decreasing downward) rather than \emph{in-situ} pressure. That is, 
\begin{equation}
\rho = \eosB(S, \theta, z) = \eos(S, \theta, -g \rB z), 
\end{equation}
 where $\rB$ is the Boussinesq reference density and $g$ the gravitational acceleration \citep{young2010}. 
The preceding theory is modified to the Boussinesq case simply by swapping $p$ for $z$ and $R$ for $B$. For instance, the \emph{in-situ} density gradient becomes
\begin{equation}
 \nabla \rho = \rho_S \nabla S + \rho_\theta \nabla \theta + \rho_z \nabla z,
\end{equation} 
where now $\rho_S = \di_S B(S,\theta,z)$, $\rho_\theta = \di_\theta B(S,\theta,z)$, and $\rho_z = \di_z B(S,\theta,z) < 0$. Also, $\nabla z = \khat$.
The neutral surface relation \eqref{eq:def_ntp_svp} becomes
\begin{equation}
\label{eq:def_ntp_rz}
 \nabla \rn = \utilde{\rho_z} \nabla \zn.
\end{equation} 
Now $\rfn$ and $\rpfn$ are functions of $z$, not $p$. However, the essence of the theory is unchanged. 
(In practice, the topobaric code trivially switches to the Boussinesq case by using $B$ instead of $R$, and internally treating $z$ as positive and increasing downwards, like $p$.)

\subsection{Results}
\label{sec:topobaric_results}

To assess topobaric surfaces, they are compared against five other approximately neutral surfaces. Computation time is briefly discussed, but mostly the comparison rests on neutrality. 

\subsubsection{Six classes of approximately neutral surfaces}

Six types of approximately neutral surfaces will be constructed:
\begin{enumerate}[(a)]
\item potential density surfaces \citep{wust1935}, isosurfaces of $\sigma_1 = \eosB(S,\theta,\SI{-1000}{m})$ or $\sigma_2 = \eosB(S,\theta,\SI{-2000}{m})$;
\item \emph{in-situ} density anomaly surfaces \citep{montgomery1937}, isosurfaces of $\delta = \rho - \eosB(S_\delta,\theta_\delta,z)$ where $S_\delta$ and $\theta_\delta$ are constants;
\item neutral density surfaces, isosurfaces of $\gamma^n$ \citep{jackett.mcdougall1997}
\item $\sigma_\nu$-surfaces, ``orthobaric'';
\item $\omega$-surfaces \citep{klocker.mcdougall.ea2009};
\item $\tau$-surfaces, topobaric. 
\end{enumerate}
The ``orthobaric'' surface is not actually an isosurface of the \citet{deszoeke.springer.ea2000} 3D orthobaric density; quotation marks around ``orthobaric'' help indicate this distinction. Rather, it is a special case of a topobaric surface but with the whole surface fit together in a single region, and $\rzfn$ fit as a cubic spline with knots at $z_{max}$, \SI{-200}{m}, \SI{-1500}{m}, \SI{-1800}{m}, and $z_{min}$, where $z_{max}$ and $z_{min}$ are the shallowest and deepest depths found on the surface, respectively. This allows us to test the importance of geography in topobaric surfaces. 
Similarly, the $\sigma_1$-, $\sigma_2$-, and $\delta$-surfaces are not constructed as isosurfaces of 3D scalar fields using vertical interpolation, but rather as solutions of a non-linear equation in each water column, found by bisection, \eg at each $\bm{x}$, solving for $z$ in $\rho(\bm{x},z) - \eosB(S_\delta,\theta_\delta,z) - \delta  = 0$ for some constant (isovalue) $\delta$.

The $\omega$-surface is constructed first; it is initialized from a $\sigma_2$-surface intersecting ($\xref$, $\yref$, \SI{-2000}{m}), but it heaves during its iterative procedure, finishing at ($\xref$, $\yref$, \SI{\zreftwo}{m}). 
The other five surfaces are constructed to intersect this latter point. Specifically, this means
(a) $\sigma_2 = 1036.9551~\mathrm{kg}~\mathrm{m}^{-3}$, 
(b) $\delta = -3.4759 \times 10^{-3}~\mathrm{kg}~\mathrm{m}^{-3}$ with $S_\delta = \SI{34.6568}{psu}$ and $\theta_\delta = 2.0899^\circ$C taken as mean values on the $\omega$-surface between $55^\circ$S and $50^\circ$S (in the Southern Ocean, the nexus of the other oceans) , 
and (c) $\gamma^n = 27.9248$.  
Moreover, the ``orthobaric'' surface (d) and the topobaric surface (f) are initialized from the isopycnal (a) with $\bm{x}_\rv$ = ($\xref$, $\yref$) and reference depth $z_\rv = \SI{\zreftwo}{m}$ (the analogue of the reference pressure $p_\rv$ for the Boussinesq case). 

\begin{figure*}[tb]
  \centering\includegraphics[width=1\textwidth]{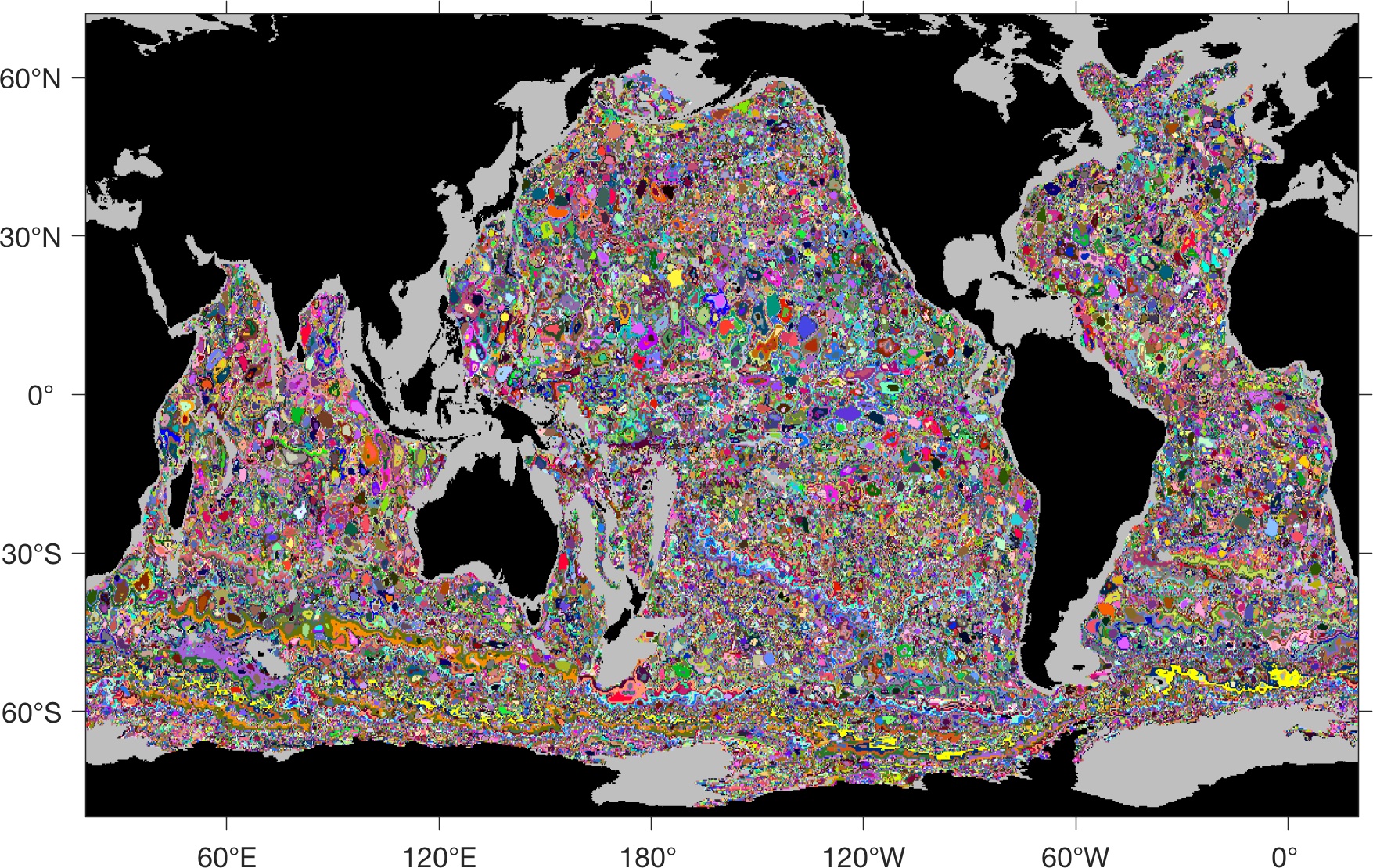}
  \caption{The associated regions (colours) for the Reeb graph of the depth of the $\sigma_2$ isopycnal intersecting ($\xref$, $\yref$, \SI{\zreftwo}{m}), which initializes the algorithm to create a topobaric surface. A single-valued function is empirically fit for each distinct region. Grey regions indicate the isopycnal has outcropped or grounded, or has disconnected from the main surface.
  }
\label{fig:argmap}
\end{figure*}

Figure~\ref{fig:argmap} shows the associated regions for the Reeb graph of $\zn$ for the $\sigma_2$-surface that initializes the topobaric surface calculation. This is part of step 2 in the first iteration to create the $\tau$-surface.
Some large eddies can be seen as individual regions, and zonally elongated structures can be seen in the Southern Ocean. As there are over 40,000 arcs, the full structure is far beyond comprehension. It is possible to simplify the graph as in Section~\ref{sec:illustrative}, but the present goal---to produce as neutral a surface as possible---is best met by keeping all the fine-scale structure of the Reeb graph.

\subsubsection{Computation time}

The topobaric surface (f) is constructed in five iterations, and the convergence is quite rapid: the root-mean-square change of $\zn$ after the first iteration is \SI{15}{m}, then \SI{90}{cm}, \SI{5.0}{cm}, \SI{2.8}{mm}, \SI{0.45}{mm}, after iterations two through five.
On a (single) 2.2~GHz processor, the $\tau$-surface was computed in under 50~seconds. Using the original code, the $\omega$-surface requires about half a day (50 iterations). Vectorizing the original MATLAB code reduced this to about 30~minutes. The $\omega$-surface code remains under development, so a precise speed comparison is not the present goal. 

Topobaric surfaces have the advantage that the Reeb graph (which is quick to compute) mostly decouples the problem. Each arc $a$ of the Reeb graph that is not on a cycle (which is the vast majority of arcs) requires fitting an affine linear function in $\rzfn_a(z)$; this involves a dense $n$ by 2 matrix, where $n$ is the number of water columns in the region $\reg_a$. Arcs on cycles are fit in a larger, coupled problem to satisfy the cycle constraint. This involves a block matrix with as many submatrices (dense $n$ by 2 matrices as above) as there are arcs in the cycle basis, arranged diagonally, plus one equality constraint for each cycle in the cycle basis. Then, updating $\zn$ is completely decoupled, being a single  problem per water column, though non-linear.

In contrast, the bottleneck for $\omega$-surfaces is finding the least-squares solution of a coupled linear equation for the entire ocean. This involves a sparse $m$ by $n$ matrix, where $n$ is the number of water columns in the rectilinearly gridded ocean and $m \approx 2n$. This matrix is banded with five non-zero entries per row, plus a row of ones at the bottom to conserve density.

\subsubsection{Neutrality}

The neutrality of the six surfaces is now compared. In fact, neutrality is one of three desirable properties of a quasi-conservative variable (such as potential temperature in a dry atmosphere); the other two are material conservation and the existence of an exact geostrophic streamfunction \citep{deszoeke.springer2009}.  Material conservation cannot be assessed because topobaric surfaces are 2D, not 3D structures. Topobaric surfaces possess an exact geostrophic streamfunction \citep{stanley2019geostrf}. Thus, attention rests on neutrality. 

The error from neutrality \eqref{eq:def_ntp_rz} is measured by 
\begin{equation}
\label{eq:epsilon}
\bm{\epsilon} = \nabla \rn - \utilde{\rho_z} \nabla \zn,
\end{equation}
which is zero for a perfectly neutral surface. 
Numerically, $\nabla \rn$ and $\nabla \zn$ are calculated by non-centred finite differences, and $\utilde{\rho_z}$ is evaluated from the equation of state using $\sn$, $\tn$, and $\zn$ averaged between the two grid points involved in the aforementioned finite difference. 
This is a third order accurate discretisation: expanding both terms in this discretisation of $\nabla \rn$ using a Taylor series about the averaged $\sn$, $\tn$, and $\zn$ reveals the quadratic terms cancel identically. 

A second measure of error, and one for which we have a more familiar numeric sense, is the diapycnal diffusivity caused by the isopycnal diffusivity when the surface is not perfectly aligned with the neutral tangent plane. This is called the fictitious diapycnal diffusivity \citep{mcdougall.jackett2005assessment, klocker.mcdougall.ea2009}, expressed as
\begin{equation}
D^f = K \, \bm{s} \cdot \bm{s}
\end{equation}
where $K$ is an isopycnal eddy diffusivity, taken as a representative constant $K = 1000~\mathrm{m}^2~\mathrm{s}^{-1}$, and 
\begin{equation}
\bm{s} = \utilde{\nabla_n z} - \nabla \zn
\end{equation}
is the slope difference between the neutral tangent plane and the approximately neutral surface. 
The slope of the neutral tangent plane, $\nabla_n z$, is a vector field in 3D space, hence the under-tilde is used to restrict it to the surface in question. 
Expressing $\bm{N} = (N_1, N_2, N_3)$ in the $(\ihat, \jhat, \khat)$ basis, an explicit formula is $\nabla_n z = -(N_1/N_3) \ihat - (N_2/N_3) \jhat$.
A more useful expression for $\bm{s}$ is found as follows. Use the standard coordinate transformation
$\nabla_n \theta = \nabla_z \theta + \di_z \theta \, \nabla_n z$ and similarly for $\nabla_n S$. Here, $\nabla_z$ is a horizontal gradient at constant depth. Multiply these by $\rho_\theta$ and $\rho_S$ respectively, then sum to cancel the neutral $\theta$ and $S$ gradients, by \eqref{eq:def_ntp_st2}. Repeat this for $\nabla \tn$ and $\nabla \sn$ (for which this cancellation is not complete). Subtracting the two results yields (dropping most under-tildes for visual clarity)
\begin{equation}
\label{eq:slope_error}
\bm{s} = 
- \left( \frac{\rho_S \nabla \sn + \rho_\theta \nabla \tn}{\rho_S \, \di_z S + \rho_\theta \, \di_z \theta} \right)
= \left( \frac{g}{\rho N^2} \right) \, \bm{\epsilon}
\end{equation} 
where $N$ is the Brunt--V\"ais\"al\"a frequency, $N^2 = -g \rho^{-1} \di_z \sigma_L$, where $\sigma_L = \eosB(S,\theta,\zn)$ is the locally referenced potential density. 
Numerically, $D^f$ requires the components of $\bm{s}$ on the same grid, so now $\bm{\epsilon}$ in \eqref{eq:slope_error} is computed using centred differences for $\nabla \rn$ and $\nabla \zn$, and $\utilde{\rho_z}$ is computed from the equation of state using $\sn$, $\tn$, and $\zn$ averaged between the two adjacent water columns (maintaining the third order accuracy), not the central column (tempting as that is). Piecewise Cubic Hermite Interpolating Polynomials (PCHIPs) are used to evaluate $\di_z \sigma_L$. 
As $D^f \propto N^{-4}$, global statistics of $D^f$ can be overwhelmed by a single grid point with $N \approx 0$, such as in mode water, so $N$ is artificially increased to a minimum value of $2 \times 10^{-4}~\mathrm{s}^{-1}$.

\begin{figure*}[!t]
  \centering
  \includegraphics[width=1\textwidth]
{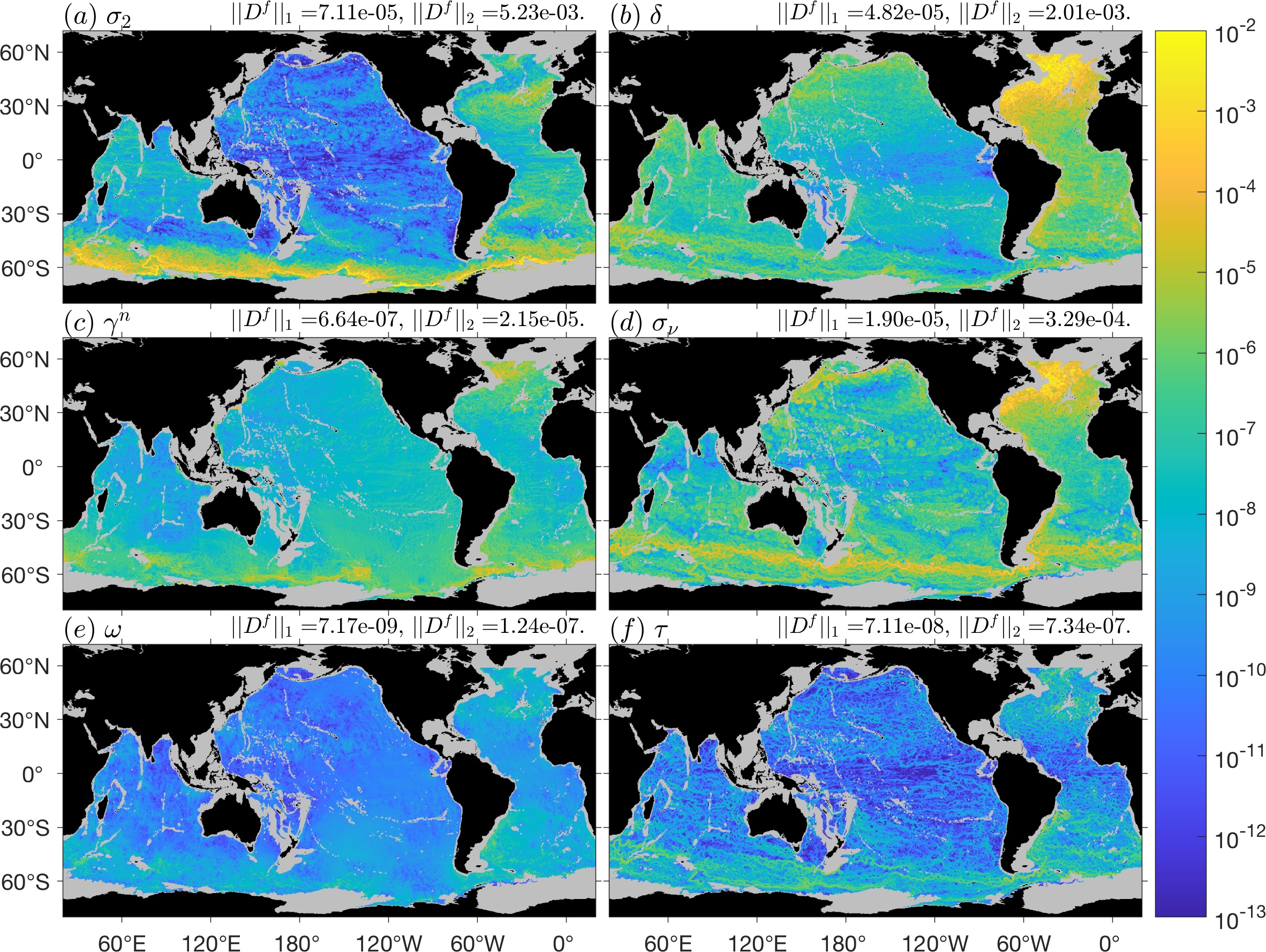}
  \caption{Comparison of the fictitious diapycnal diffusivity, $D^f$, on six approximately neutral surfaces, all intersecting ($\xref$, $\yref$, \SI{\zreftwo}{m}): 
(a) an isosurface of potential density referenced to \SI{-2000}{m}, 
(b) an isosurface of \emph{in-situ} density anomaly referenced to $34.6568$~psu, and $2.0899^\circ$C, and
(c) an isosurface of neutral density, 
(d) an ``orthobaric'' surface,
(e) an $\omega$-surface, and
(f) a topobaric surface. 
  Surfaces on the right (left) do (not) possess an exact geostrophic streamfunction. 
A common mask (grey) is applied to all regions. This mask is the largest connected region from those points that are valid on all six surfaces (note neutral density is not defined north of $64^\circ$N, and $\omega$-surfaces exclude the mixed layer). 
  The area-weighted $l_1$ and $l_2$ norms of these errors are listed above each panel.
  }
\label{fig:fddmaps}
\end{figure*}

\begin{figure*}[!t]
  \centering\includegraphics[width=1\textwidth]{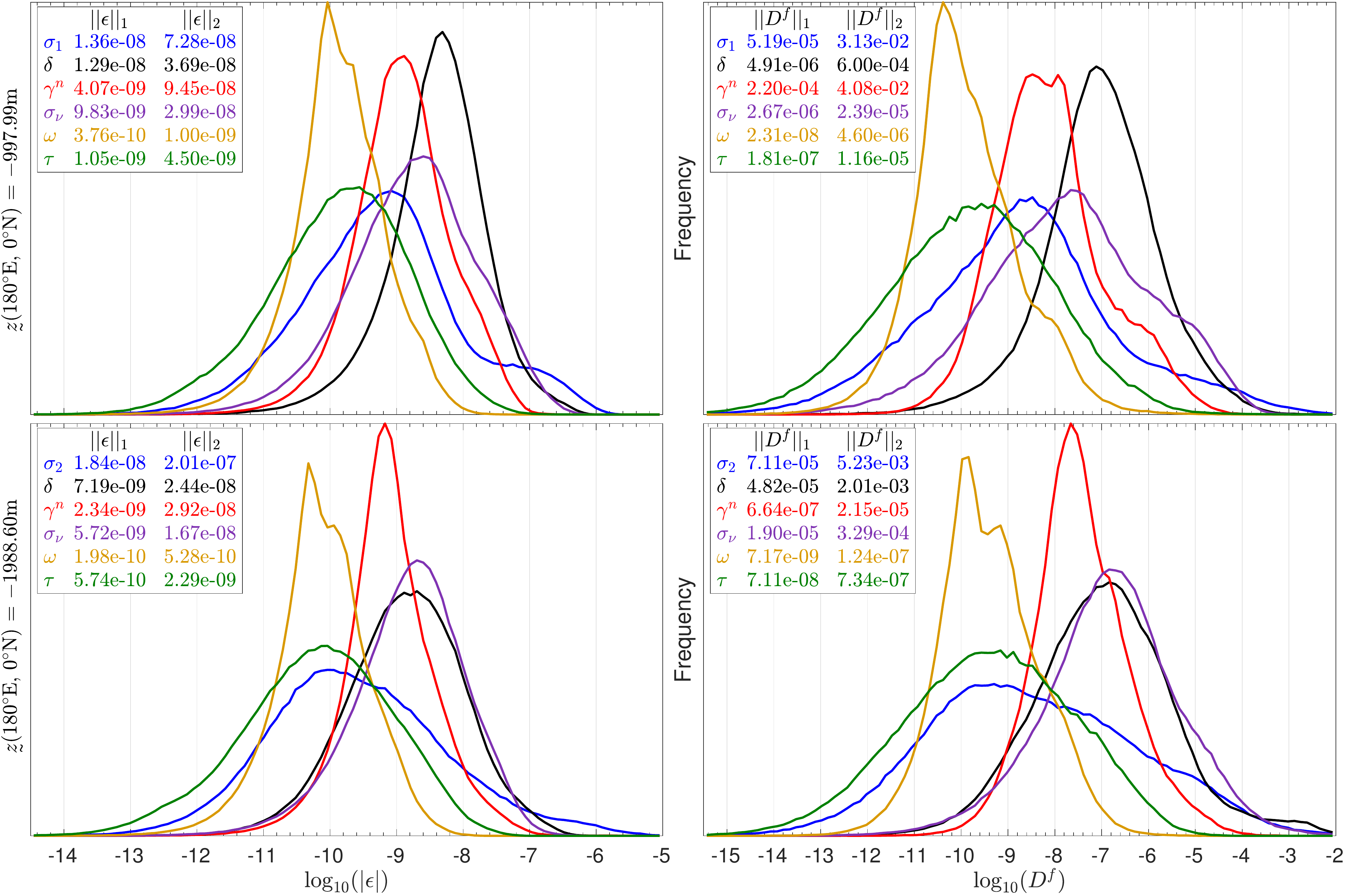}
  \caption{Logarithmic histograms of the zonal and meridional components of the neutrality error $\bm{\epsilon}$ (left, units kg~m$^{-4}$ dropped), and the fictitious diapycnal diffusivity $D^f$ (right, units m$^2$~s$^{-1}$ dropped), for six surfaces all at depth \SI{\zrefone}{m} (top) or \SI{\zreftwo}{m} (bottom) at ($\xref$, $\yref$).
  Also listed are the area-weighted $l_1$ and $l_2$ norms, each calculated over a common geographic region (see text for details).
  }
\label{fig:hist6}
\end{figure*}

Figure~\ref{fig:fddmaps} maps $D^f$ for these six surfaces, with a common mask applied so $D^f$ is shown only where it is valid on all six surfaces. 
The $\sigma_2$-surface performs well (low $D^f$) where it is near the reference depth of \SI{-2000}{m}, such as in the Pacific and Indian Oceans, and to a lesser extent in the Atlantic Ocean; however, it performs very poorly in the Southern Ocean where it rises to the sea-surface. 
In contrast, the $\delta$-surface performs decently in the Southern Ocean where the salinity and potential temperature are close to their reference values, but poorly elsewhere, particularly in the North Atlantic. 
The reference values for these two surfaces can always be chosen so they perform well in a limited geographic region, but they struggle globally. 
The $\gamma^n$-surface overcomes this problem, yielding reasonably low $D^f$ globally. 
The underlying $4^\circ$ by $4^\circ$ grid of the WOCE atlas from which neutral density interpolates \citep{jackett.mcdougall1997} does produce artefacts, as can be seen in the Southern Ocean where $D^f$ changes sharply. 
The $\sigma_\nu$-surface exhibits high $D^f$ in both the Southern Ocean and North Atlantic, but not so high as for the $\sigma_2$- or $\delta$-surfaces in these regions. 
Both the $\omega$-surface and $\tau$-surface exhibit very small $D^f$ globally, and both exhibit their largest $D^f$ in the Southern Ocean and North Atlantic, right where neutral helicity is largest \citep[][Fig.~4c]{klocker.mcdougall.ea2009}. 

Whereas the $\omega$-surface spreads $D^f$ errors quite smoothly over the global ocean, the $\tau$-surface exhibits much more eddy-scale, filamentary structure in $D^f$. 
Some similar structure is visible on the $\sigma_2$-surface in the Pacific, where it is extremely neutral, suggesting this structure is partly real.
However, it may also be caused by neighbouring grid points that are in regions associated with different arcs. 
This is a numerical difficulty arising from the finite difference underlying the calculation of $\bm{\epsilon}$: in the limit as the grid spacing goes to zero, $\bm{\epsilon}$ would not compare grid points across regions (except on contours through $\pn$ saddles).
This issue is exacerbated when neighbouring grid points are in regions associated with arcs that are not incident to a common node in the Reeb graph, as can happen for filamentary regions. 
The topobaric algorithm could be conceivably modified to smooth these numerical errors out, for instance by inflating each associated region by one grid point in every direction when fitting $\rzfn$. This has not been tested, but even without such algorithmic enhancements, the $\tau$-surface performs very well. 

For a more quantitative comparison, Fig.~\ref{fig:hist6} shows histograms of $\bm{\epsilon}$ 
and $D^f$ for the above six surfaces, and for a further set of these six surfaces that intersect ($\xref$, $\yref$, \SI{\zrefone}{m})\footnote{
Again the $\omega$-surface is calculated first, initialized from a $\sigma_1$-surface intersecting ($\xref$, $\yref$, \SI{-1000}{m}). 
For the other surfaces, 
(a) $\sigma_1 = 1032.0053~\mathrm{kg}~\mathrm{m}^{-3}$, 
(b) $\delta =   -1.4026 \times 10^{-2}~\mathrm{kg}~\mathrm{m}^{-3}$ 
with $S_\delta = \SI{34.3389}{psu}$ 
and $\theta_\delta = 3.0272^\circ$C, 
(c) $\gamma^n =   27.5400$, 
(d) the ``orthobaric'' surface uses a cubic spline for $\rzfn$ with knots only at 0, -200, and \SI{-6000}{m},
and (d,f) the ``orthobaric'' and topobaric surfaces have reference location $\bm{x}_\rv$ = ($\xref$, $\yref$) and reference depth $z_\rv = \SI{\zrefone}{m}$, and are initialized from the isopycnal (a). 
}. 
The area-weighted $l_1$ (mean absolute error) and $l_2$ (root mean square error) norms are listed as inset tables in Fig.~\ref{fig:hist6}. 
The data underlying each histogram and norm is best thought of as a 1D array. For $D_f$, this is an array of all $D_f$ values at the grid cell centres within a common mask where $D_f$ is valid on all six surfaces intersecting a common depth at ($\xref$, $\yref$). 
For $\bm{\epsilon}$, this array is roughly twice as long, containing the zonal component of $\bm{\epsilon}$ within a similarly constructed common mask, followed by the meridional component of $\bm{\epsilon}$ within a similarly constructed common mask. 

First consider the surfaces at roughly \SI{-2000}{m} in the Pacific (bottom row).
Compared to the $\sigma_2$-surface, the $\gamma^n$-surface reduces the metrics $||\bm{\epsilon}||_1$, $||\bm{\epsilon}||_2$, $||D^f||_1$, and $||D^f||_2$ by factors of 
7.9, 6.9, 107, and 243, respectively.
Whereas $||\bm{\epsilon}||_1$ involves a sum of zonal and meridional components of $\bm{\epsilon}$, $||D^f||_1$ involves a sum of $ \bm{\epsilon} \cdot \bm{\epsilon} = \epsilon^2$, and hence punishes surfaces with more heterogeneous errors. Indeed, the $\sigma_2$-surface's poor performance in the Southern Ocean is particularly crippling to its $D^f$ metrics. 
This punishment becomes even more severe for $||D^f||_2$, which involves sums of ${\epsilon}^4$. 
The weighting $g / (\rho N^2)$ actually lowers $D^f$ for the $\sigma_2$-surface: the stratification $N^2$ is larger in the Southern Ocean as these surfaces rise towards the sea-surface (they tend to avoid low $N^2$ mode water, which occupies little space in density coordinates).
The $\tau$-surface further improves upon the $\gamma^n$-surface, reducing $||D^f||_2$, for instance, by a factor of 29.
Also, the $\tau$-surface, with its geographic dependence, vastly outperforms its geographically independent cousin, the $\sigma_\nu$-surface: $||D^f||_2$ differs by a factor of 448.
Still, the $\omega$-surface performs best, with $||\epsilon||_2$ and $||D^f||_2$ smaller by factors of 4.3 and 5.9, respectively, than those of the $\tau$-surface.

\sloppy This analysis does not assess which of neutral density and  orthobaric density is superior, a subject of some debate \citep{mcdougall.jackett2005assessment,deszoeke.springer2009}. 
Indeed, ``orthobaric'' surfaces are computed here essentially from the topobaric algorithm, rather than from the algorithm (and dataset) of \citet{deszoeke.springer.ea2000}. Moreover, material conservation has not been evaluated.
Topobaric surfaces, though, take the best of both worlds: 
they contain geographic dependence which enables neutral density to be more neutral, while retaining the good theoretical properties of orthobaric density. 

Ideally, the fictitious diapycnal diffusivity is less than the true diapycnal diffusivity. Taking a representative value of $10^{-5}~\mathrm{m}^2~\mathrm{s}^{-1}$ for the latter, the fraction of ocean over which this is false, for surfaces (a) through (f), is 
4.9\%, 7.7\%, 1.0\%, 8.5\%, $2.3 \times 10^{-5}$, and $8.5 \times 10^{-4}$, respectively.
 Indeed, the histograms of Fig.~\ref{fig:hist6} have a long tail towards high $D^f$ for the $\sigma_2$- and $\delta$- and $\sigma_\nu$-surfaces. The area where $D^f$ exceeds this threshold is very small for both $\omega$- and $\tau$-surfaces. 

The comparison between the $\omega$- and $\tau$-surfaces is similar on the surfaces around \SI{-1000}{m} in the Pacific, with 
$||\bm{\epsilon}||_1$, $||\bm{\epsilon}||_2$, $||D^f||_1$, and $||D^f||_2$ smaller for the $\omega$-surface by factors of 2.8, 4.5, 7.8, and 2.5, respectively.
In terms of these metrics, the $\gamma^n$ surface performs worst at this depth, but this is due to a block of troublesome points south of Java, where the depth of the $\gamma^n$-surface changes by over \SI{-1000}{m} between neighbouring grid points.  This unrealistic behaviour is caused by the neutral density software's underlying $4^\circ \times 4^\circ$ grid, which must have had different bathymetry.
The histograms in Fig.~\ref{fig:hist6} reveal the $\gamma^n$-surface, aside from these points, again outperforms the $\sigma_1$-, $\delta$-, and $\sigma_\nu$-surfaces.
In general, neutral density performs somewhat worse than it could because the \citet{levitus1982} climatology underlying neutral density differs from the ECCO2 state, and neutral density uses an older equation of state \citep{millero.chen.ea1980}.

As $\omega$-surfaces minimize $||\bm{\epsilon}||_2$, there can be no hope\footnote{
Actually, $\omega$-surfaces minimize $||\bm{\epsilon}||_2$ without changing its two-dimensional curl in the surface. 
This approach rests on an approximate relationship derived by \citet{mcdougall.jackett1988} that relates the 2D curl of $\bm{\epsilon}$ in an approximately neutral surface to $H N^{-2}$. 
To the extent that this approximation is good, the 2D curl of $\bm{\epsilon}$ is set by the ocean hydrography, hence $\omega$-surfaces do not attempt to change it: the \citet{klocker.mcdougall.ea2009} algorithm takes an initial surface with an initial $\bm{\epsilon}$, and finds $\Phi'$ to minimize $||\bm{\epsilon} + \nabla \Phi'||_2$.  
The quality of this approximation determines the degree to which methods that vary the 2D curl of $\bm{\epsilon}$ might discover surfaces with smaller $||\bm{\epsilon}||_2$.
}
 to produce an approximately neutral surface with smaller $||\bm{\epsilon}||_2$.
Nonetheless, topobaric surfaces are quick to compute and perform admirably, while also also possessing an exact geostrophic streamfunction \citep{stanley2019geostrf}.

\section{Conclusions}
\label{sec:conclusions}

\sloppy The pathways along which the ocean, below the mixed layer, is connected are largely determined by neutral surfaces. The typical oceanic epineutral diffusivity is $\mathcal{O}(10^3~\mathrm{m}^2~\mathrm{s}^{-1})$ while dianeutral diffusivity is $\mathcal{O}(10^{-5}~\mathrm{m}^2~\mathrm{s}^{-1})$ in the main thermocline \citep{mackinnon.stlaurent.ea2013}. It is therefore crucial to orient the large epineutral diffusion correctly, in the neutral tangent plane, lest a component of it act dianeutrally and swamp the real dianeutral diffusion. This can be quantified as a fictitious dianeutral diffusivity \citep{mcdougall.jackett2005material, klocker.mcdougall.ea2009}.
Relatedly, the lateral velocity that acts in the neutral tangent plane must pierce any approximately neutral surface, creating a flow across that surface. \citet{klocker.mcdougall2010} quantified this flow, finding upwards of \SI{10}{Sv} globally through deep $\sigma_0$-surfaces. This diapycnal flow is entirely fictitious, due to $\sigma_0$-surfaces' poor alignment with the neutral tangent plane. They find this is reduced to generally less than \SI{1}{Sv} when using $\omega$-surfaces.

The neutral tangent plane is parallel to a surface of locally referenced potential density \citep{mcdougall1987ns}. Depth-level ocean models long ago rotated their diffusion tensors \citep{veronis1975, redi1982} to align with the neutral tangent plane. This is fairly straightforward because it is a local problem---the neutral tangent plane is well-defined essentially everywhere. However, layered models require a quasi-conservative density variable for their vertical coordinate, which is a global problem, and one that is not well-defined: neutral tangent planes cannot be globally stitched together to form a well-defined neutral surface, because of non-zero neutral helicity \citep{mcdougall.jackett1988}. 
As such, well-defined surfaces can only be approximately neutral.
Many oceanic theories and analyses operate on neutral surfaces, or would do so if neutral surfaces were well-defined surfaces with a unique depth. We have shown that this is not guaranteed even in an ocean with zero neutral helicity, as neutral helices can exist around islands and other holes in a neutral surface.

Nowadays, a host of approximately neutral surfaces are available.
Many of them are purely thermodynamic variables (functions only of salinity, temperature, and pressure), such as
potential density \citep{wust1935}, 
specific volume anomaly \citep{montgomery1937},
orthobaric density \citep{deszoeke.springer.ea2000},
a rational approximation of neutral density \citep{mcdougall.jackett2005material},
and 
thermodynamic neutral density \citep{tailleux2016generalized}.
Global isosurfaces of these variables are limited in their neutrality not so much by the non-zero helicity of the real ocean, but overwhelmingly by the fact that they lack any geographic dependence. 
That is, even in an ocean with zero neutral helicity, a conservative density variable must be a function of latitude and longitude, as well as salinity, temperature, and pressure. 
Neutral density \citep{jackett.mcdougall1997} is such a function,  explicitly containing geographic information.
Geographic dependence is implicitly built into $\omega$-surfaces \citep{klocker.mcdougall.ea2009} by the rectilinear grid on which it operates. As such, $\gamma^n$- and $\omega$-surfaces can be close to neutral, globally. 

The importance of geography arises because there is a multivalued functional relationship between \emph{in-situ} density and pressure (equivalently, between practical/Absolute salinity and potential/Conservative temperature) on a neutral surface. 
This multivalued function has single-valued branches within certain geographic regions, and these branches differ between regions. 
This was well-known, but the shape (topology) of these geographic regions was unknown. 

Much emphasis has been placed on these functional relationships differing between the Northern and Southern hemispheres \citep{mcdougall.jackett2005assessment}.
\citet{deszoeke.springer2005} advanced the original orthobaric density to use a different virtual compressibility for the North and the South Atlantic. However, the result is discontinuous at the equator, and a fictitious force must be applied to a water parcel crossing this discontinuity to keep it on the same orthobaric density surface \citep{deszoeke.springer2005}.
Following this idea to its limit, \citet{deszoeke.springer2009} developed ``extended orthobaric density'' by segmenting the Atlantic into arbitrarily many latitude bins; this minimizes the discontinuities, but creates more of them. 
Somewhat similarly, patched potential density \citep{reid.lynn1971} and generalized patched potential density \citep{tailleux2016generalized} add geographic dependence to potential density by segmenting the ocean into boxes aligned with latitude circles, longitude circles, and depth or pressure levels. 
However, these are not the correct geographic shapes underpinning the multivalued functional relationship between \emph{in-situ} density and pressure on neutral surfaces. 

The cause of the geographic dependence may also have been unknown. 
\citet{mcdougall.jackett2005assessment} correctly stated that, on a neutral surface having pressure $\pn$, new branches of the functional relationship between \emph{in-situ} density and pressure open up at points where $\nabla \pn = \bm{0}$. However, thinking in terms of neutral trajectories in a zonally uniform ocean, they stated that this occurs at extrema of $\pn$. 
Extrema of $\pn$ do have $\nabla \pn = \bm{0}$, but it is more appropriate to say these points \emph{close} branches of the multivalued function, and branches open at saddle points of $\pn$.
(Branches often exist that do not enclose any extrema, such as the red $AB$ arc in Fig.~\ref{fig:schematic}. A better example is the cycle in Fig.~\ref{fig:schematic}, bearing no relation at all to $\pn$ extrema. If $\pn$ has one maxima and one minima and $n$ islands, there can be up to $1+3n$ branches.) 

The first major advance of this paper is to reveal the entire geographic structure underlying this multivalued function, by use of the Reeb graph of $\pn$. 
Each such region is mapped to an arc of the Reeb graph, and nodes represent the critical points (saddles and extrema) of $\pn$.
The structure of the graph---which arcs are incident to which nodes---determines how the geographic regions nest into a global structure. 
Relationships between branches of the multivalued function are determined by the structure of the graph.

Topobaric surfaces represent the second major advance of this paper. Knowing the geographic regions underlying different branches of the multivalued relation between \emph{in-situ} density and pressure, these branches can be fit empirically, subject to some matching conditions related to cycles in the Reeb graph. A root-finding problem ensues, solving for the pressure in each water column for which the \emph{in-situ} density at that pressure exactly matches the single-valued function at that pressure. In this way, an iterative procedure turns an isopycnal (or any approximately neutral surface) into a topobaric surface. Topobaric surfaces have an exact multivalued functional relation between \emph{in-situ} density and pressure, are very close to neutral, and possess an exact geostrophic streamfunction \citep{stanley2019geostrf}.

Topobaric surfaces are the topologically correct extension of isosurfaces of orthobaric density to have geographic dependence. 
Orthobaric density is pycnotropic---a function only of pressure and \emph{in-situ} density. It employs a pycnotropic virtual compressibility that approximates the real compressibility.
On an isosurface of orthobaric density, the \emph{in-situ} density is an implicit function of pressure, so the virtual compressibility is a function only of pressure, analogous to how topobaric surfaces approximate the real compressibility\footnote{We call $\rho_p$ the compressibility, rather $\rho^{-1} \rho_p$, a convention also used by \citet{deszoeke.springer.ea2000}.}, $\rpn$, by $\rpfn(\pn)$ using a multivalued function $\rpfn$. 
The discontinuities of extended orthobaric density \citep{deszoeke.springer2009} are caused by changing the virtual compressibility at latitudinal boundaries. No such discontinuities exist for topobaric surfaces, because the branches of $\rpfn$ change at contours of constant $\pn$. Future work aims to develop a 3D topobaric density variable. 

This topological musing bears on the Lorenz convention, which sets the depth of a surface that has outcropped or incropped to be infinitesimally below the sea-surface or above the sea-floor \citep{young2012}. In this sense, the only holes in the surface are islands---quite a simplification. Is this justifiable? From the topological perspective, two contours of the same level set become one contour that snakes around the hole; points on either side of the hole on the same level set are glued together in topological space. But the fundamental reason for alignment of contours of salinity and potential temperature on a neutral surface is because both tracers are materially advected by the flow, of which there is none through a hole. The Lorenz convention makes sense from the perspective of one fluid column, considering also the density of air and solid Earth, but not from a broader perspective. For example, different water masses exist on different sides of submarine ridges (and so on to smaller features), and it would be wrong to join them.

Using the Reeb graph as a (computational) tool to study multivalued functions arising from equations like \eqref{eq:def_ntp_svp} is a new endeavour. 
This is highly translatable to other problems.
A multivalued functional relationship between variables is a powerful idea, so fresh insights on other problems may soon be discovered by similar means. 

\section*{Acknowledgements}
The author thanks Chris Hughes and David Marshall for helpful discussions, 
Trevor McDougall for early encouragement on related work that ultimately led to this paper,
two reviewers who helped clarify this paper,
Andreas Klocker for sharing his $\omega$-surface software,
David Gleich for his graph theory MATLAB toolbox (GAIMC), 
and Harish Doraiswamy and Vijay Natarajan for their ReCon software. 
The author was supported by the Clarendon Scholarship, and the Canadian Alumni Scholarship at Linacre College, University of Oxford. 
MATLAB software to compute topobaric surfaces is available from the author's website.

\appendix

\section{Graph theory glossary}
\label{sec:mathdefs_graphtheory}
A \textdef{graph} is a tuple $G = (N, A)$, where
$N$ is a set of \textdef{nodes}
and $A$ is a set of \textdef{arcs}, and elements of $A$ are subsets of $N$ with precisely two elements. 
Intuitively, a graph is a collection of nodes, and arcs connecting any two nodes. 

When $A$ is a multiset (its elements need not be unique), $G$ is a \textdef{multigraph}. A multigraph can have multiple arcs incident upon the same pair of nodes.  
The Reeb graph is actually a multigraph, but ``graph'' is used as shorthand.

Arc $a$ is \textdef{incident upon} node $n$ when $n \in a$. 

Two nodes are \textdef{adjacent} if there is an arc incident upon both nodes. 

Two arcs are \textdef{incident} if they are both incident upon a common node.

The \textdef{degree} of a node is the number of distinct arcs incident upon it.

A \textdef{walk} is an ordered sequence that alternates between nodes and arcs such that every arc is incident upon both nodes next to it in the sequence. In standard graph theory, a walk must start and end with nodes, but we relax that here. 

A \textdef{(simple) cycle} is a walk whose first node is also its last node, and otherwise contains no repeated vertices, and contains no repeated arcs. 

A \textdef{connected} graph is a graph such that there is a walk between any two of its nodes.

A \textdef{tree} is a connected graph with no cycles.

A \textdef{spanning subgraph} of a graph $G = (N,A)$ is a graph $H = (N, F)$ with $F \subseteq A$. 

A \textdef{minimum spanning tree} is a spanning subgraph that is also a tree. 

An \textdef{Eulerian graph} is a graph whose nodes all have even degree. 

The \textdef{cycle space} of a graph $G$ is the set of all Eulerian spanning subgraphs of $G$. 
Two elements of the cycle space may be added by symmetric difference of their arc sets to produce a new element of the cycle space. 
A \textdef{cycle basis} is a set of simple cycles, a subset of which can be combined by symmetric difference on their arc sets to produce any element of the cycle space.

\section{Neutral helix pitch around islands}
\label{sec:island_pitch}

\begin{figure}[!t]
\centering
\includegraphics[width=\columnwidth]{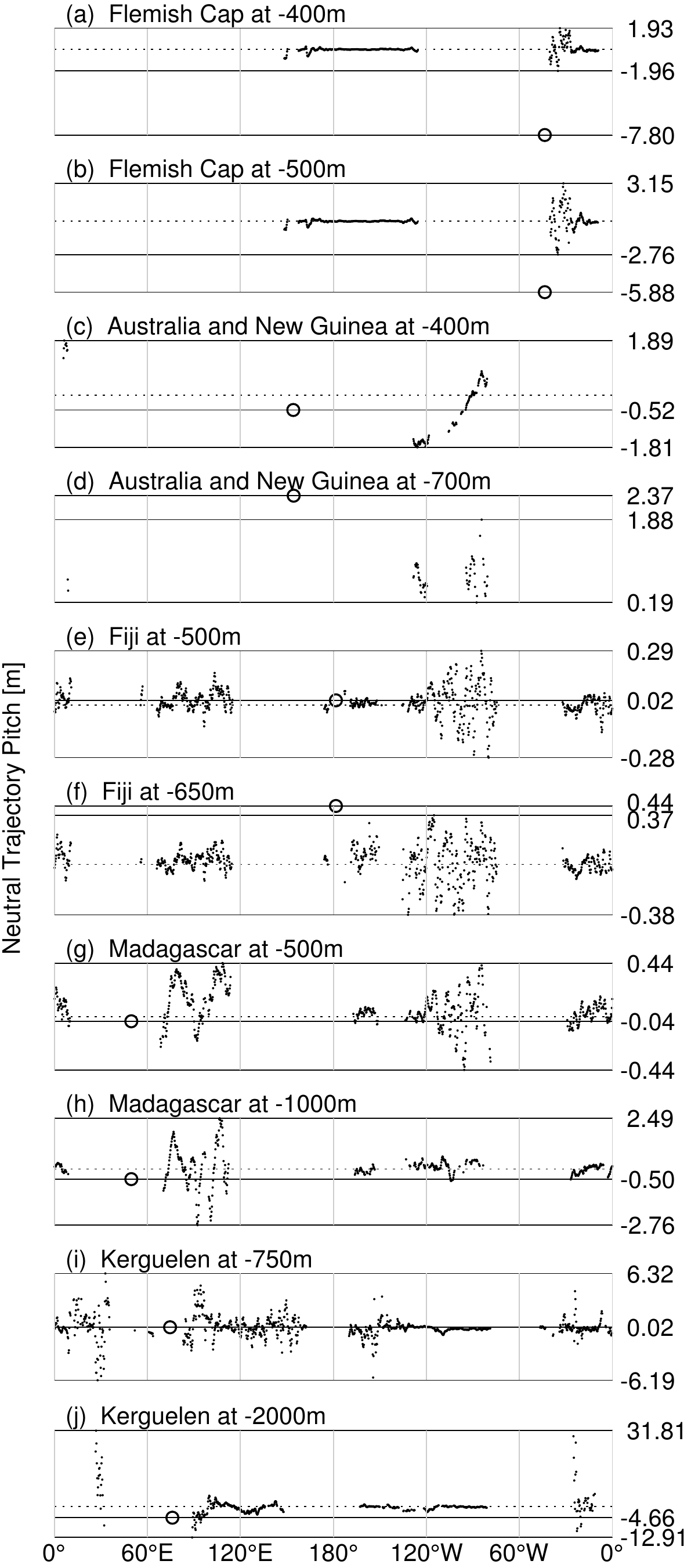}
\caption{
The pitch of a neutral helix that adaptively traces the boundary of an island/seamount, initialized 1 grid point east of the island at a specified depth (as indicated above each panel), shown by an open circle. 
Dots show the pitch of neutral helices initialized at the same depth and having the same shape as that around the island, but shifted to an arbitrary longitude (abscissa); gaps indicate where that neutral trajectory grounded or outcropped. 
Horizontal lines indicate the maximum and minimum pitch of all such trajectories, and the pitch around the island (indicated at right, in metres); the dashed line indicates zero pitch.
}
\label{fig:island_pitch}
\end{figure}

A neutral trajectory that returns to its starting water column does so, in general, at a different depth from which it began. Is this change---the pitch of a neutral helix---fundamentally different depending on whether the enclosed area is open ocean, or an island/seamount?
A preliminary analysis towards this question is shown here, using ECCO2 data \citep{menemenlis.hill.ea2005} from 22--24 December 2002. 

Given one water column and an initial depth $z_A$, a neutral trajectory to a (nearby) water column ends at depth $z_B$ if 
\begin{equation}
\label{eq:neutral_traj}
\rho(S_A(\overline{z}), \theta_A(\overline{z}), \overline{z})
- \rho(S_B(\overline{z}), \theta_B(\overline{z}), \overline{z}) = 0,
\end{equation}
where $\overline{z} = (z_A + z_B) / 2$, and where $S_i(z)$ and $\theta_i(z)$ are the salinity and potential temperature as functions of depth in water column $i \in \{A,B\}$ \citep{jackett.mcdougall1997}.  The functions are taken as piecewise linear interpolants, and \eqref{eq:neutral_traj} is solved by bisection, geometrically expanding outward from an initial guess of $z_A$ until a sign change is found. 

A neutral trajectory around an island/seamount is discovered by starting just east of an island at a particular depth, then making neutral trajectories one grid cell at a time. The forward direction is initially west. A neutral trajectory in the forward direction is tested; if this fails (no solution, meaning the neutral trajectory grounds or outcrops) the forward direction is rotated $90^\circ$ clockwise. This is repeated until a neutral trajectory in the forward direction succeeds. That step is made, and the forward direction is rotated $90^\circ$ counter-clockwise. The whole procedure stops when the trajectory returns to the starting water column (or an upper bound of steps is reached, in which case the neutral trajectory has spiralled up or down the topographic slope, and this island must be tested at another initial depth).  

Figure~\ref{fig:island_pitch}(a) shows that a neutral helix around Flemish Cap, starting at \SI{-400}{m}, descends by \SI{7.8}{m} in one counter-clockwise loop. This is compared to other neutral helices starting at \SI{-400}{m} but at other longitudes and having the same horizontal shape (and thus the same area) as that around Flemish Cap. Most of these open-ocean neutral helices have a pitch very close to \SI{0}{m}; the maximum is \SI{1.93}{m}. Thus, it seems that Flemish Cap is exceptional amongst duplicates of Flemish Cap's shape in the open ocean. 

Figure~\ref{fig:island_pitch}(b)---(j) show the results of this analysis repeated at other islands/seamounts and starting depths. 
Moving just \SI{100}{m} deeper, Flemish Cap again appears to be exceptional.
Australia and New Guinea at \SI{-700}{m} appears to be exceptional, but not so much so at \SI{-400}{m}: a not-insignificant pitch of \SI{-0.52}{m} still exists, but this is smaller than for neutral helices of the same shape at other longitudes, which is not hard given their large area.
Fiji at \SI{-650}{m} appears to be exceptional, but not so at \SI{-500}{m}.
The neutral helix pitch around Madagascar at \SI{-500}{m} and around Kerguelen at \SI{-750}{m} are both tiny. Around Madagascar at \SI{-1000}{m} the neutral helix pitch is larger than equivalent helices' pitches in the Pacific and Atlantic Oceans, but not in the Indian Ocean. Neutral helices around Kerguelen starting at \SI{-2000}{m} appear to have a moderately exceptional pitch. 

Not all islands/seamounts should necessarily be exceptional in this way---only those which produce a cycle in the Reeb graph of $\zn$. A sense of which islands/seamounts create cycles is gained by looking at depth contours on a potential density surface referenced to the starting depth of the neutral trajectory (not shown). Flemish Cap, Australia and New Guinea, and Fiji should clearly create cycles, because $\zn$ contours are found emanating from them and intersecting land elsewhere. 
This is also true of Madagascar at \SI{-1000}{m} and Kerguelen at \SI{-2000}{m}, around both of which the neutral helix pitch is large. However, for Madagascar at \SI{-500}{m}, a $\zn = \SI{-450}{m}$ contour very nearly encircles Madagascar---it barely intersects the northern end of Madagascar, skims the coast of Africa intersecting it only slightly in four places, and just intersects the southern tip of Nazareth Bank to the east of Madagascar. There are a few small seamounts contained within this $\zn = \SI{-450}{m}$ contour, which can create cycles around Madagascar. Still, there is a hint here that Madagascar should not give rise to the path-dependency responsible for neutral helices, which would explain why the pitch of a neutral helix around Madagascar at \SI{-500}{m} is so small, only \SI{-4}{cm}. Similarly, for Kerguelen at \SI{-750}{m}, the $\zn = \SI{-500}{m}$ and $\zn = \SI{-700}{m}$ contours that flank Kerguelen happen to traverse the entire Southern Ocean without intersecting land (or nearly so---the northern one does skim five small seamounts, which collectively occupy only 35 grid points). Again, this is suggestive that neutral helices around Kerguelen at \SI{-750}{m} should have a pitch near zero, in agreement the numerical calculation of a pitch of only \SI{2}{cm}. However, it is not clear why the neutral helix around Fiji starting at \SI{-500}{m} is so small, as Fiji should create a cycle in the Reeb graph.

The depths, and to some extent the islands, in Fig.~\ref{fig:island_pitch} were chosen with some care, to illustrate exceptional and unexceptional cases. The actual results seem fairly sensitive to the initial location. 
For example, the pitch of neutral helices around Madagascar as a function of initial depth is a fairly complicated function; it is generally increasing with depth, but has six \SI{0}{m} crossings shallower than \SI{-1250}{m} and seven \SI{1}{m} crossings between \SI{-1250}{m} and \SI{-2250}{m} (not shown).

Are islands/seamounts---in particular those that intersect depth contours of approximately neutral surfaces so as to create cycles in the Reeb graph---exceptional at producing neutral helices with large pitches?  Further work is needed to definitively say, but this preliminary analysis suggests it is a possibility. 

\section*{References}

\bibliography{stanley_bibtex}

\begin{thebibliography}{45}
\expandafter\ifx\csname natexlab\endcsname\relax\def\natexlab#1{#1}\fi
\providecommand{\url}[1]{\texttt{#1}}
\providecommand{\href}[2]{#2}
\providecommand{\path}[1]{#1}
\providecommand{\DOIprefix}{doi:}
\providecommand{\ArXivprefix}{arXiv:}
\providecommand{\URLprefix}{URL: }
\providecommand{\Pubmedprefix}{pmid:}
\providecommand{\doi}[1]{\href{http://dx.doi.org/#1}{\path{#1}}}
\providecommand{\Pubmed}[1]{\href{pmid:#1}{\path{#1}}}
\providecommand{\bibinfo}[2]{#2}
\ifx\xfnm\relax \def\xfnm[#1]{\unskip,\space#1}\fi
\bibitem[{Arnol'd(1957)}]{arnold1957}
\bibinfo{author}{Arnol'd, V.I.}, \bibinfo{year}{1957}.
\newblock \bibinfo{title}{On the representability of a function of two
  variables in the form $\chi[\phi(x)+\psi(y)]$}.
\newblock \bibinfo{journal}{Uspekhi Matematicheskikh Nauk}
  \bibinfo{volume}{12}, \bibinfo{pages}{119--121}.
\bibitem[{Arnold(2006)}]{arnold2006}
\bibinfo{author}{Arnold, V.I.}, \bibinfo{year}{2006}.
\newblock \bibinfo{title}{From {{Hilbert}}'s superposition problem to dynamical
  systems}, in: \bibinfo{booktitle}{Mathematical Events of the Twentieth
  Century}. \bibinfo{publisher}{{Springer}}, pp. \bibinfo{pages}{19--47}.
\bibitem[{Biasotti et~al.(2008)Biasotti, Giorgi, Spagnuolo and
  Falcidieno}]{biasotti.giorgi.ea2008}
\bibinfo{author}{Biasotti, S.}, \bibinfo{author}{Giorgi, D.},
  \bibinfo{author}{Spagnuolo, M.}, \bibinfo{author}{Falcidieno, B.},
  \bibinfo{year}{2008}.
\newblock \bibinfo{title}{Reeb graphs for shape analysis and applications}.
\newblock \bibinfo{journal}{Theoretical Computer Science}
  \bibinfo{volume}{392}, \bibinfo{pages}{5--22}.
\newblock \DOIprefix\doi{10.1016/j.tcs.2007.10.018}.
\bibitem[{Carr et~al.(2006)Carr, Moller and Snoeyink}]{carr.moller.ea2006}
\bibinfo{author}{Carr, H.}, \bibinfo{author}{Moller, T.},
  \bibinfo{author}{Snoeyink, J.}, \bibinfo{year}{2006}.
\newblock \bibinfo{title}{Artifacts caused by simplicial subdivision}.
\newblock \bibinfo{journal}{IEEE Transactions on Visualization and Computer
  Graphics} \bibinfo{volume}{12}, \bibinfo{pages}{231--242}.
\newblock \DOIprefix\doi{10.1109/TVCG.2006.22}.
\bibitem[{Carr et~al.(2003)Carr, Snoeyink and Axen}]{carr.snoeyink.ea2003}
\bibinfo{author}{Carr, H.}, \bibinfo{author}{Snoeyink, J.},
  \bibinfo{author}{Axen, U.}, \bibinfo{year}{2003}.
\newblock \bibinfo{title}{Computing contour trees in all dimensions}.
\newblock \bibinfo{journal}{Computational Geometry} \bibinfo{volume}{24},
  \bibinfo{pages}{75--94}.
\newblock \DOIprefix\doi{10.1016/S0925-7721(02)00093-7}.
\bibitem[{Carr et~al.(2010)Carr, Snoeyink and {van de
  Panne}}]{carr.snoeyink.ea2010}
\bibinfo{author}{Carr, H.}, \bibinfo{author}{Snoeyink, J.},
  \bibinfo{author}{{van de Panne}, M.}, \bibinfo{year}{2010}.
\newblock \bibinfo{title}{Flexible isosurfaces: {{Simplifying}} and displaying
  scalar topology using the contour tree}.
\newblock \bibinfo{journal}{Computational Geometry} \bibinfo{volume}{43},
  \bibinfo{pages}{42--58}.
\newblock \DOIprefix\doi{10.1016/j.comgeo.2006.05.009}.
\bibitem[{{Cole-McLaughlin} et~al.(2003){Cole-McLaughlin}, Edelsbrunner, Harer,
  Natarajan and Pascucci}]{cole-mclaughlin.edelsbrunner.ea2003}
\bibinfo{author}{{Cole-McLaughlin}, K.}, \bibinfo{author}{Edelsbrunner, H.},
  \bibinfo{author}{Harer, J.}, \bibinfo{author}{Natarajan, V.},
  \bibinfo{author}{Pascucci, V.}, \bibinfo{year}{2003}.
\newblock \bibinfo{title}{Loops in reeb graphs of 2-manifolds},
  \bibinfo{publisher}{{ACM Press}}. p. \bibinfo{pages}{344}.
\newblock \DOIprefix\doi{10.1145/777792.777844}.
\bibitem[{{de Szoeke} and Springer(2005)}]{deszoeke.springer2005}
\bibinfo{author}{{de Szoeke}, R.A.}, \bibinfo{author}{Springer, S.R.},
  \bibinfo{year}{2005}.
\newblock \bibinfo{title}{The all-{{Atlantic}} temperature-salinity-pressure
  relation and patched potential density}.
\newblock \bibinfo{journal}{Journal of Marine Research} \bibinfo{volume}{63},
  \bibinfo{pages}{59--93}.
\newblock \DOIprefix\doi{10.1357/0022240053693752}.
\bibitem[{{de Szoeke} and Springer(2009)}]{deszoeke.springer2009}
\bibinfo{author}{{de Szoeke}, R.A.}, \bibinfo{author}{Springer, S.R.},
  \bibinfo{year}{2009}.
\newblock \bibinfo{title}{The {{Materiality}} and {{Neutrality}} of {{Neutral
  Density}} and {{Orthobaric Density}}}.
\newblock \bibinfo{journal}{Journal of Physical Oceanography}
  \bibinfo{volume}{39}, \bibinfo{pages}{1779--1799}.
\newblock \DOIprefix\doi{10.1175/2009JPO4042.1}.
\bibitem[{{de Szoeke} et~al.(2000){de Szoeke}, Springer and
  Oxilia}]{deszoeke.springer.ea2000}
\bibinfo{author}{{de Szoeke}, R.A.}, \bibinfo{author}{Springer, S.R.},
  \bibinfo{author}{Oxilia, D.M.}, \bibinfo{year}{2000}.
\newblock \bibinfo{title}{Orthobaric density: {{A}} thermodynamic variable for
  ocean circulation studies}.
\newblock \bibinfo{journal}{Journal of physical oceanography}
  \bibinfo{volume}{30}, \bibinfo{pages}{2830--2852}.
\bibitem[{Doraiswamy and Natarajan(2013)}]{doraiswamy.natarajan2013}
\bibinfo{author}{Doraiswamy, H.}, \bibinfo{author}{Natarajan, V.},
  \bibinfo{year}{2013}.
\newblock \bibinfo{title}{Computing {{Reeb Graphs}} as a {{Union}} of {{Contour
  Trees}}}.
\newblock \bibinfo{journal}{IEEE Transactions on Visualization and Computer
  Graphics} \bibinfo{volume}{19}, \bibinfo{pages}{249--262}.
\newblock \DOIprefix\doi{10.1109/TVCG.2012.115}.
\bibitem[{Forget(2010)}]{forget2010}
\bibinfo{author}{Forget, G.}, \bibinfo{year}{2010}.
\newblock \bibinfo{title}{Mapping {{Ocean Observations}} in a {{Dynamical
  Framework}}: {{A}} 2004\textendash{}06 {{Ocean Atlas}}}.
\newblock \bibinfo{journal}{Journal of Physical Oceanography}
  \bibinfo{volume}{40}, \bibinfo{pages}{1201--1221}.
\newblock \DOIprefix\doi{10.1175/2009JPO4043.1}.
\bibitem[{Heine et~al.(2011)Heine, Schneider, Carr and
  Scheuermann}]{heine.schneider.ea2011}
\bibinfo{author}{Heine, C.}, \bibinfo{author}{Schneider, D.},
  \bibinfo{author}{Carr, H.}, \bibinfo{author}{Scheuermann, G.},
  \bibinfo{year}{2011}.
\newblock \bibinfo{title}{Drawing {{Contour Trees}} in the {{Plane}}}.
\newblock \bibinfo{journal}{IEEE Transactions on Visualization and Computer
  Graphics} \bibinfo{volume}{17}, \bibinfo{pages}{1599--1611}.
\newblock \DOIprefix\doi{10.1109/TVCG.2010.270}.
\bibitem[{Iselin(1939)}]{iselin1939}
\bibinfo{author}{Iselin, C.O.}, \bibinfo{year}{1939}.
\newblock \bibinfo{title}{The influence of vertical and lateral turbulence on
  the characteristics of the waters at mid-depths}.
\newblock \bibinfo{journal}{Transactions, American Geophysical Union}
  \bibinfo{volume}{20}, \bibinfo{pages}{414}.
\newblock \DOIprefix\doi{10.1029/TR020i003p00414}.
\bibitem[{Jackett and Mcdougall(1995)}]{jackett.mcdougall1995}
\bibinfo{author}{Jackett, D.R.}, \bibinfo{author}{Mcdougall, T.J.},
  \bibinfo{year}{1995}.
\newblock \bibinfo{title}{Minimal {{Adjustment}} of {{Hydrographic Profiles}}
  to {{Achieve Static Stability}}}.
\newblock \bibinfo{journal}{Journal of Atmospheric and Oceanic Technology}
  \bibinfo{volume}{12}, \bibinfo{pages}{381--389}.
\newblock \DOIprefix\doi{10.1175/1520-0426(1995)012<0381:MAOHPT>2.0.CO;2}.
\bibitem[{Jackett and McDougall(1997)}]{jackett.mcdougall1997}
\bibinfo{author}{Jackett, D.R.}, \bibinfo{author}{McDougall, T.J.},
  \bibinfo{year}{1997}.
\newblock \bibinfo{title}{A neutral density variable for the world's oceans}.
\newblock \bibinfo{journal}{Journal of Physical Oceanography}
  \bibinfo{volume}{27}, \bibinfo{pages}{237--263}.
\newblock \DOIprefix\doi{10.1175/1520-0485(1997)027<0237:ANDVFT>2.0.CO;2}.
\bibitem[{Klocker and McDougall(2010)}]{klocker.mcdougall2010}
\bibinfo{author}{Klocker, A.}, \bibinfo{author}{McDougall, T.J.},
  \bibinfo{year}{2010}.
\newblock \bibinfo{title}{Influence of the {{Nonlinear Equation}} of {{State}}
  on {{Global Estimates}} of {{Dianeutral Advection}} and {{Diffusion}}}.
\newblock \bibinfo{journal}{Journal of Physical Oceanography}
  \bibinfo{volume}{40}, \bibinfo{pages}{1690--1709}.
\newblock \DOIprefix\doi{10.1175/2010JPO4303.1}.
\bibitem[{Klocker et~al.(2009)Klocker, McDougall and
  Jackett}]{klocker.mcdougall.ea2009}
\bibinfo{author}{Klocker, A.}, \bibinfo{author}{McDougall, T.J.},
  \bibinfo{author}{Jackett, D.R.}, \bibinfo{year}{2009}.
\newblock \bibinfo{title}{A new method for forming approximately neutral
  surfaces}.
\newblock \bibinfo{journal}{Ocean Science} \bibinfo{volume}{5},
  \bibinfo{pages}{155--172}.
\newblock \DOIprefix\doi{10.5194/os-5-155-2009}.
\bibitem[{Levitus(1982)}]{levitus1982}
\bibinfo{author}{Levitus, S.}, \bibinfo{year}{1982}.
\newblock \bibinfo{title}{Climatological atlas of the world ocean}.
\newblock \bibinfo{journal}{NOAA Profess. Pap.} \bibinfo{volume}{13},
  \bibinfo{pages}{1--173}.
\bibitem[{Lynn and Reid(1968)}]{lynn.reid1968}
\bibinfo{author}{Lynn, R.J.}, \bibinfo{author}{Reid, J.L.},
  \bibinfo{year}{1968}.
\newblock \bibinfo{title}{Characteristics and circulation of deep and abyssal
  waters}.
\newblock \bibinfo{journal}{Deep Sea Research and Oceanographic Abstracts}
  \bibinfo{volume}{15}, \bibinfo{pages}{577--598}.
\newblock \DOIprefix\doi{10.1016/0011-7471(68)90064-8}.
\bibitem[{MacKinnon et~al.(2013)MacKinnon, St~Laurent and
  Naveira~Garabato}]{mackinnon.stlaurent.ea2013}
\bibinfo{author}{MacKinnon, J.}, \bibinfo{author}{St~Laurent, L.},
  \bibinfo{author}{Naveira~Garabato, A.C.}, \bibinfo{year}{2013}.
\newblock \bibinfo{title}{Diapycnal {{Mixing Processes}} in the {{Ocean
  Interior}}}, in: \bibinfo{booktitle}{International {{Geophysics}}}.
  \bibinfo{publisher}{{Elsevier}}. volume \bibinfo{volume}{103}, pp.
  \bibinfo{pages}{159--183}.
\newblock \DOIprefix\doi{10.1016/B978-0-12-391851-2.00007-6}.
\bibitem[{McDougall(1987a)}]{mcdougall1987ns}
\bibinfo{author}{McDougall, T.J.}, \bibinfo{year}{1987}a.
\newblock \bibinfo{title}{Neutral {{Surfaces}}}.
\newblock \bibinfo{journal}{Journal of Physical Oceanography}
  \DOIprefix\doi{10.1175/1520-0485(1987)017<1950:NS>2.0.CO;2}.
\bibitem[{McDougall(1987b)}]{mcdougall1987tb}
\bibinfo{author}{McDougall, T.J.}, \bibinfo{year}{1987}b.
\newblock \bibinfo{title}{Thermobaricity, cabbeling, and water-mass
  conversion}.
\newblock \bibinfo{journal}{Journal of Geophysical Research}
  \bibinfo{volume}{92}, \bibinfo{pages}{5448}.
\newblock \DOIprefix\doi{10.1029/JC092iC05p05448}.
\bibitem[{McDougall et~al.(2014)McDougall, Groeskamp and
  Griffies}]{mcdougall.groeskamp.ea2014}
\bibinfo{author}{McDougall, T.J.}, \bibinfo{author}{Groeskamp, S.},
  \bibinfo{author}{Griffies, S.M.}, \bibinfo{year}{2014}.
\newblock \bibinfo{title}{On {{Geometrical Aspects}} of {{Interior Ocean
  Mixing}}}.
\newblock \bibinfo{journal}{Journal of Physical Oceanography}
  \bibinfo{volume}{44}, \bibinfo{pages}{2164--2175}.
\newblock \DOIprefix\doi{10.1175/JPO-D-13-0270.1}.
\bibitem[{McDougall and Jackett(1988)}]{mcdougall.jackett1988}
\bibinfo{author}{McDougall, T.J.}, \bibinfo{author}{Jackett, D.R.},
  \bibinfo{year}{1988}.
\newblock \bibinfo{title}{On the helical nature of neutral trajectories in the
  ocean}.
\newblock \bibinfo{journal}{Progress in Oceanography} \bibinfo{volume}{20},
  \bibinfo{pages}{153--183}.
\newblock \DOIprefix\doi{10.1016/0079-6611(88)90001-8}.
\bibitem[{McDougall and Jackett(2005a)}]{mcdougall.jackett2005assessment}
\bibinfo{author}{McDougall, T.J.}, \bibinfo{author}{Jackett, D.R.},
  \bibinfo{year}{2005}a.
\newblock \bibinfo{title}{An assessment of orthobaric density in the global
  ocean}.
\newblock \bibinfo{journal}{Journal of Physical Oceanography}
  \bibinfo{volume}{35}, \bibinfo{pages}{2054--2075}.
\newblock \DOIprefix\doi{10.1175/JPO2796.1}.
\bibitem[{McDougall and Jackett(2005b)}]{mcdougall.jackett2005material}
\bibinfo{author}{McDougall, T.J.}, \bibinfo{author}{Jackett, D.R.},
  \bibinfo{year}{2005}b.
\newblock \bibinfo{title}{The material derivative of neutral density}.
\newblock \bibinfo{journal}{Journal of Marine Research} \bibinfo{volume}{63},
  \bibinfo{pages}{159--185}.
\newblock \DOIprefix\doi{10.1357/0022240053693734}.
\bibitem[{McDougall and Jackett(2007)}]{mcdougall.jackett2007}
\bibinfo{author}{McDougall, T.J.}, \bibinfo{author}{Jackett, D.R.},
  \bibinfo{year}{2007}.
\newblock \bibinfo{title}{The thinness of the ocean in $s-\theta-p$ space and
  the implications for mean diapycnal advection}.
\newblock \bibinfo{journal}{Journal of Physical Oceanography}
  \bibinfo{volume}{37}, \bibinfo{pages}{1714--1732}.
\newblock \DOIprefix\doi{10.1175/JPO3114.1}.
\bibitem[{Menemenlis et~al.(2005)Menemenlis, Hill, Adcrocft, Campin, Cheng,
  Ciotti, Fukumori, Heimbach, Henze, K\"ohl, Lee, Stammer, Taft and
  Zhang}]{menemenlis.hill.ea2005}
\bibinfo{author}{Menemenlis, D.}, \bibinfo{author}{Hill, C.},
  \bibinfo{author}{Adcrocft, A.}, \bibinfo{author}{Campin, J.M.},
  \bibinfo{author}{Cheng, B.}, \bibinfo{author}{Ciotti, B.},
  \bibinfo{author}{Fukumori, I.}, \bibinfo{author}{Heimbach, P.},
  \bibinfo{author}{Henze, C.}, \bibinfo{author}{K\"ohl, A.},
  \bibinfo{author}{Lee, T.}, \bibinfo{author}{Stammer, D.},
  \bibinfo{author}{Taft, J.}, \bibinfo{author}{Zhang, J.},
  \bibinfo{year}{2005}.
\newblock \bibinfo{title}{{{NASA}} supercomputer improves prospects for ocean
  climate research}.
\newblock \bibinfo{journal}{Eos, Transactions American Geophysical Union}
  \bibinfo{volume}{86}, \bibinfo{pages}{89}.
\newblock \DOIprefix\doi{10.1029/2005EO090002}.
\bibitem[{Millero et~al.(1980)Millero, Chen, Bradshaw and
  Schleicher}]{millero.chen.ea1980}
\bibinfo{author}{Millero, F.J.}, \bibinfo{author}{Chen, C.T.},
  \bibinfo{author}{Bradshaw, A.}, \bibinfo{author}{Schleicher, K.},
  \bibinfo{year}{1980}.
\newblock \bibinfo{title}{A new high pressure equation of state for seawater}.
\newblock \bibinfo{journal}{Deep Sea Research Part A. Oceanographic Research
  Papers} \bibinfo{volume}{27}, \bibinfo{pages}{255--264}.
\newblock \DOIprefix\doi{10.1016/0198-0149(80)90016-3}.
\bibitem[{Montgomery(1937)}]{montgomery1937}
\bibinfo{author}{Montgomery, R.}, \bibinfo{year}{1937}.
\newblock \bibinfo{title}{A suggested method for representing gradient flow in
  isentropic surfaces}.
\newblock \bibinfo{journal}{Bull. Amer. Meteor. Soc} \bibinfo{volume}{18},
  \bibinfo{pages}{210--212}.
\bibitem[{Nycander(2011)}]{nycander2011}
\bibinfo{author}{Nycander, J.}, \bibinfo{year}{2011}.
\newblock \bibinfo{title}{Energy {{Conversion}}, {{Mixing Energy}}, and
  {{Neutral Surfaces}} with a {{Nonlinear Equation}} of {{State}}}.
\newblock \bibinfo{journal}{Journal of Physical Oceanography}
  \bibinfo{volume}{41}, \bibinfo{pages}{28--41}.
\newblock \DOIprefix\doi{10.1175/2010JPO4250.1}.
\bibitem[{Redi(1982)}]{redi1982}
\bibinfo{author}{Redi, M.H.}, \bibinfo{year}{1982}.
\newblock \bibinfo{title}{Oceanic {{Isopycnal Mixing}} by {{Coordinate
  Rotation}}}.
\newblock \bibinfo{journal}{Journal of Physical Oceanography}
  \bibinfo{volume}{12}, \bibinfo{pages}{1154--1158}.
\newblock \DOIprefix\doi{10.1175/1520-0485(1982)012<1154:OIMBCR>2.0.CO;2}.
\bibitem[{Reeb(1946)}]{reeb1946}
\bibinfo{author}{Reeb, G.}, \bibinfo{year}{1946}.
\newblock \bibinfo{title}{Sur les points singuliers d'une forme de {{Pfaff}}
  completement int\'egrable ou d'une fonction num\'erique}.
\newblock \bibinfo{journal}{CR Acad. Sci. Paris} \bibinfo{volume}{222},
  \bibinfo{pages}{2}.
\bibitem[{Reid and Lynn(1971)}]{reid.lynn1971}
\bibinfo{author}{Reid, J.L.}, \bibinfo{author}{Lynn, R.J.},
  \bibinfo{year}{1971}.
\newblock \bibinfo{title}{On the influence of the {{Norwegian}}-{{Greenland}}
  and {{Weddell}} seas upon the bottom waters of the {{Indian}} and {{Pacific}}
  oceans}.
\newblock \bibinfo{journal}{Deep Sea Research and Oceanographic Abstracts}
  \bibinfo{volume}{18}, \bibinfo{pages}{1063--1088}.
\newblock \DOIprefix\doi{10.1016/0011-7471(71)90094-5}.
\bibitem[{Sneddon(1957)}]{sneddon1957}
\bibinfo{author}{Sneddon, I.}, \bibinfo{year}{1957}.
\newblock \bibinfo{title}{Elements of Partial Differential Equations}.
\newblock International Series in Pure and Applied Mathematics,
  \bibinfo{publisher}{{McGraw-Hill}}.
\bibitem[{Stanley(2018)}]{stanley2018thesis}
\bibinfo{author}{Stanley, G.}, \bibinfo{year}{2018}.
\newblock \bibinfo{title}{Tales from {{Topological Oceans}}}.
\newblock Ph.D. thesis. University of Oxford.
\bibitem[{Stanley(2019)}]{stanley2019geostrf}
\bibinfo{author}{Stanley, G.J.}, \bibinfo{year}{2019}.
\newblock \bibinfo{title}{The exact geostrophic streamfunction for neutral
  surfaces}.
\newblock \bibinfo{journal}{Ocean Modelling} .
\bibitem[{Starr(1945)}]{starr1945}
\bibinfo{author}{Starr, V.P.}, \bibinfo{year}{1945}.
\newblock \bibinfo{title}{A {{Quasi}}-{{Lagrangian System}} of {{Hydrodynamical
  Equations}}}.
\newblock \bibinfo{journal}{Journal of Meteorology} \bibinfo{volume}{2},
  \bibinfo{pages}{227--237}.
\newblock \DOIprefix\doi{10.1175/1520-0469(1945)002<0227:AQLSOH>2.0.CO;2}.
\bibitem[{Tailleux(2016)}]{tailleux2016generalized}
\bibinfo{author}{Tailleux, R.}, \bibinfo{year}{2016}.
\newblock \bibinfo{title}{Generalized {{Patched Potential Density}} and
  {{Thermodynamic Neutral Density}}: {{Two New Physically Based
  Quasi}}-{{Neutral Density Variables}} for {{Ocean Water Masses Analyses}} and
  {{Circulation Studies}}}.
\newblock \bibinfo{journal}{Journal of Physical Oceanography}
  \bibinfo{volume}{46}, \bibinfo{pages}{3571--3584}.
\newblock \DOIprefix\doi{10.1175/JPO-D-16-0072.1}.
\bibitem[{Tarjan(1972)}]{tarjan1972}
\bibinfo{author}{Tarjan, R.}, \bibinfo{year}{1972}.
\newblock \bibinfo{title}{Depth-{{First Search}} and {{Linear Graph
  Algorithms}}}.
\newblock \bibinfo{journal}{SIAM Journal on Computing} \bibinfo{volume}{1},
  \bibinfo{pages}{146--160}.
\newblock \DOIprefix\doi{10.1137/0201010}.
\bibitem[{Veronis(1975)}]{veronis1975}
\bibinfo{author}{Veronis, G.}, \bibinfo{year}{1975}.
\newblock \bibinfo{title}{The role of models in tracer studies}.
\newblock \bibinfo{journal}{Numerical models of ocean circulation} ,
  \bibinfo{pages}{133--146}.
\bibitem[{W\"ust(1935)}]{wust1935}
\bibinfo{author}{W\"ust, G.}, \bibinfo{year}{1935}.
\newblock \bibinfo{title}{The stratosphere of the {{Atlantic}} ocean}.
\newblock \bibinfo{journal}{Scientific Results of the German Atlantic
  Expedition of the Research Vessel ``Meteor'' 1925\textendash{}27}
  \bibinfo{volume}{6}.
\bibitem[{Young(2010)}]{young2010}
\bibinfo{author}{Young, W.R.}, \bibinfo{year}{2010}.
\newblock \bibinfo{title}{Dynamic {{Enthalpy}}, {{Conservative Temperature}},
  and the {{Seawater Boussinesq Approximation}}}.
\newblock \bibinfo{journal}{Journal of Physical Oceanography}
  \bibinfo{volume}{40}, \bibinfo{pages}{394--400}.
\newblock \DOIprefix\doi{10.1175/2009JPO4294.1}.
\bibitem[{Young(2012)}]{young2012}
\bibinfo{author}{Young, W.R.}, \bibinfo{year}{2012}.
\newblock \bibinfo{title}{An {{Exact Thickness}}-{{Weighted Average
  Formulation}} of the {{Boussinesq Equations}}}.
\newblock \bibinfo{journal}{Journal of Physical Oceanography}
  \bibinfo{volume}{42}, \bibinfo{pages}{692--707}.
\newblock \DOIprefix\doi{10.1175/JPO-D-11-0102.1}.

\end{thebibliography}

\end{document}